\documentclass[11pt,a4paper]{article}
\usepackage{jheppub}
\usepackage{amsmath}
\usepackage{graphicx}
\usepackage{slashed}
\usepackage[utf8]{inputenc}
\usepackage[caption=false]{subfig}

\def\eps{\epsilon}

\allowdisplaybreaks[3]

\makeatletter
\renewcommand\@fpheader{} %\hfill \parbox{3cm}{MITP/14-??\\ZU-TH~??/14}
\renewcommand\@journal{}
\makeatother

\title{Pentagon functions for massless planar scattering amplitudes}

\preprint{{ZU-TH 25/18, MPP-2018-180,  MITP/18-067, IPPP/18/59}}

\author[a]{T.~Gehrmann,}
\author[b,c]{J.M.~Henn,}
\author[d]{N.A.~Lo~Presti}

\affiliation[a]{
  Physik-Institut, 
  Universit\"at Z\"urich, Wintherturerstrasse 190,
  CH-8057~Z\"urich, Switzerland}

  \affiliation[b]{
  PRISMA Cluster of Excellence, Institute of Physics,
  Johannes Gutenberg University,\\
  D-55099~Mainz, Germany}

  \affiliation[c]{
  MPI f\"ur Physik, 
  Werner-Heisenberg-Institut,\\
  M\"unchen, Germany}

  \affiliation[d]{
  Institute for Particle Physics Phenomenology, Durham University,
  Durham DH1 3LE, UK}  

\emailAdd{thomas.gehrmann@uzh.ch}
\emailAdd{henn@uni-mainz.de}
\emailAdd{nicola.a.lo-presti@durham.ac.uk}


\abstract{Loop amplitudes for massless five particle scattering processes contain Feynman integrals depending on the 
external momentum invariants: pentagon functions. We perform a detailed study of the analyticity properties and cut structure of 
these functions up to two loops in the planar case, where we 
classify and identify the minimal set of basis functions. They are computed from the
 canonical form of their differential equations and expressed in terms of generalized polylogarithms, or alternatively as 
 one-dimensional integrals. We present analytical expressions and 
 numerical evaluation routines for these pentagon 
 functions, in all kinematical configurations relevant to five-particle scattering processes.}

\begin{document}
\unitlength1cm
\maketitle

\section{Introduction}
\label{sec:intro}

Predictions for scattering process in elementary particle physics can be computed to high accuracy through a 
perturbation series expansion of the relevant scattering amplitudes. In this expansion, higher order corrections imply 
an increasing number of closed particle loops, leading to Feynman integrals over the loop momenta. While one-loop 
corrections are known for processes with arbitrary multiplicity~\cite{Ellis:2011cr}, 
two-loop corrections to scattering amplitudes have 
up to now only been derived on a case-by-case basis, mostly for two-to-two scattering processes. In going to 
higher multiplicities for two-loop amplitudes, one faces two challenges: to express the large number of Feynman 
integrals in terms of a smaller set of basis integrals (for a minimal basis, these are often called master integrals) 
and to efficiently compute these basis integrals. 

In the recent past, important progress has been made on integral reduction techniques for two-loop multi-particle
processes, both in terms of semi-analytical approaches~\cite{Ita:2015tya,Abreu:2017idw,Badger:2017jhb} as 
well as in optimizing algebraic reductions to master 
integrals~\cite{Gluza:2010ws,Schabinger:2011dz,vonManteuffel:2014ixa,Larsen:2015ped,Peraro:2016wsq,Kosower:2018obg} 
to cope with the high complexity of processes with 
five external particles~\cite{Boehm:2018fpv,Chawdhry:2018awn} and beyond. 
As a result, expressions for two-loop five-gluon amplitudes in terms of a set of basis integrals were derived, 
first for specific helicity configurations~\cite{Badger:2013gxa,Gehrmann:2015bfy,Badger:2015lda,Dunbar:2016aux}, 
and most recently for the general-helicity case~\cite{Badger:2017jhb,Abreu:2017hqn}.

These basis integrals can be expressed in terms of a set of master 
integrals~\cite{Gehrmann:2015bfy,Papadopoulos:2015jft}: massless two-loop five-point functions. A subset of these 
functions are two-loop four-point functions with one off-shell leg, which were computed 
in analytical form~\cite{Gehrmann:2000zt,Gehrmann:2001ck,Gehrmann:2002zr}
already long ago in the context of lower multiplicity processes. The genuine five-point master integrals can be 
separated into planar and non-planar topologies, depending on the internal momentum routing. For the planar 
integrals, differential equations in the external momentum invariants~\cite{Gehrmann:1999as,Henn:2013pwa} 
were derived and 
solved in~\cite{Gehrmann:2015bfy} and~\cite{Papadopoulos:2015jft}. In this paper, we build upon our 
work on the planar master integrals in~\cite{Gehrmann:2015bfy} by taking a systematic approach aiming to 
combine the differential equations with knowledge on the kinematical analyticity structure of the master integrals 
to identify the minimal set of functions that can appear in them. These {\it pentagon functions} are the basic 
building blocks for two-loop five point master integrals, and our procedure used for their identification 
can be expected to generalise to non-planar integrals, to higher multiplicities and to higher loop orders. 
Various representations for these  pentagon functions can be derived using the differential equation method. 

Fully analytical expressions are found in terms of generalized polylogarithms~\cite{Remiddi:1999ew,Goncharov:2001iea} or 
Chen iterated integrals~\cite{Chen:1977oja}. 
These functions have a long history in the mathematics of Feynman diagrams. In recent years, huge progress was made in understanding how to handle these functions systematically.
This concerns in particular the multi-variable case. 
An important tool is the so-called `symbol' of iterated integrals \cite{Goncharov:2010jf,iterated1,2001math3059G,Duhr:2011zq}, which makes it easy to understand identities between different functions.
Closely related to this is the idea of defining the special functions needed for Feynman integrals from certain canonical differential equations \cite{Henn:2013pwa}. The latter encode the relevant data about the functions in a compact and unique way.  In particular, they contain the `symbol alphabet', denoting the integration kernels that are allowed to appear in the iterated integrals.

While the fully analytical expressions enable detailed studies of analyticity properties and of asymptotic properties, 
their numerical evaluation is rather inefficient. Instead, following \cite{Caron-Huot:2014lda},
one can derive one-dimensional integral representations, that are optimised for numerical integration.

The paper is structured as follows. Following a brief description of the kinematics of five-particle scattering in 
Euclidean and Minkowskian 
space in Section~\ref{sec:kin} and of the notation used for the planar two-loop five-point master 
integrals in Section~\ref{sec:mi}, we derive and analyse the differential equations for these master integrals in
 Section~\ref{sec:differentialequations}. Confronting the singularity structure of the differential equations 
 with the physical requirements on the analyticity properties of the master integrals provides strong 
 constraints on the pentagon functions that are allowed in the solutions of the differential equations, as 
 discussed in Section~\ref{sec:pentagonfunctions}. The pentagon functions are then computed in 
 Section~\ref{sec:boundary} by matching the generic solutions of the differential equations onto 
 boundary conditions at specific kinematical points. In Sections~\ref{sec:results} and 
\ref{sec:numerical-package}, we describe a variety of consistency checks on these results and introduce 
a public numerical code which evaluates the pentagon functions and master integrals. 
We conclude with an outlook in Section~\ref{sec:conc}.

\section{Kinematics}
\label{sec:kin}
\subsection{Conventions and Lorentz invariants}

The kinematics is described by five external momenta, $p_i^\mu$, subject to the on-shell conditions $p_{i}^2=0$, and momentum conservation $\sum_{i=1}^{5} p_{i}^\mu =0 $.
From the momenta, we can build ten scalar products $s_{ij} = 2 p_{i} \cdot p_{j}$, of which five are independent. 

We choose the following five
\begin{align}
v_1 = s_{12}/\mu^2 \,,\qquad  v_2 = s_{23}/\mu^2 \,,\qquad v_3 = s_{34}/\mu^2 \,,\qquad v_4 = s_{45}/\mu^2 \,,\qquad v_5 = s_{51}/\mu^2 \,,
\end{align}
as independent. 
We normalized them by an arbitrary scale $\mu^2>0$, in order to have dimensionless variables, while preserving manifest cyclic symmetry.
The scale $\mu^2$ occurs in a natural way in dimensional regularization.
The non-andjacent invariants can be written in terms
of adjacent ones as
\begin{eqnarray*}
  s_{13}\,=\,s_{45}-s_{12}-s_{23} \;\; & & \;\;
  s_{24}\,=\,s_{15}-s_{23}-s_{34}  \;\;\;\;\;\;\;\;\;
  s_{35}\,=\,s_{12}-s_{34}-s_{45}  \\
  s_{14}\,=\,s_{23}-s_{45}-s_{15} \;\; & &  \;\;
  s_{25}\,=\,s_{34}-s_{15}-s_{12}   \\
\end{eqnarray*}
Note that, in practice, we can always fix the overall scale of a quantity by dimensional arguments, so that effectively we need to deal with four-variable functions only. 
For most of the discussion, however, we prefer to keep working with the five variables $v_{i}$, as they allow to see symmetries in an easier way.
The above variables are parity even.
There is also a parity odd invariant, 
\begin{align}
\epsilon(1234) = 4 i \epsilon_{\mu \nu \rho \sigma } p_1^{\mu} p_2^{\nu} p_3^{\rho} p_4^{\sigma} \,. 
\end{align}
We can also write this as $\epsilon(1234)  = {\rm tr}( \gamma_{5} \slashed{p}_{1}  \slashed{p}_{2}  \slashed{p}_{3} \slashed{p}_{4} )$.
Note that $ \epsilon_{\mu \nu \rho \sigma }$ 
 is a four-dimensional object. 
The calculations in this paper are done for general dimension $D=4-2 \epsilon$.
However, since only four of the external momenta are independent, we can assume without loss of generality that they lie in some four-dimensional subspace.

We also introduce the following dimensionless 
 Gram determinant of the four linearly independent vectors $p_{i}$, $i=1\ldots 4$,
\begin{align}\label{defDelta}
\Delta = | 2 p_{i} \cdot p_{j} | / \mu^8 = (v_1 v_2 + v_2 v_3 -v_3 v_4 + v_4 v_5 -v_5 v_1)^2-4 v_1 v_2 v_3 (v_2 -v_4 -v_5) \,. 
\end{align}
It is related to the parity-odd Lorentz invariant $\epsilon(1234)$ according to 
\begin{align}
\mu^8\, \Delta =  \epsilon(1234)^2\,. \end{align}

\subsection{Physical region}

The physical regions of the five-point functions are given by $2\to 3$ scattering kinematics $(i+j \to k+l+m)$. Any pair of two 
momenta $(i,j)$ can be incoming, such that the physical region in Minkowski space corresponds to 
ten distinct regions (channels), which are commonly labelled by their initial-state invariant as $s_{ij}$-channel.
For each channel, there are constraints on the signs of the the kinematic
invariants defined as the scalar products between 
two external momenta.
These are summarised in Table~\ref{tab:channels}. 
\begin{table}[t]
\begin{center}
\begin{tabular}{ |c|c|c|c|c|c| } 
 \hline
  & \shortstack{\\In\\\,\\\,\,} & \shortstack{\\Out\\\,\\\,\,}  & \shortstack{\\Adjacent\\invariants\\($s_{12},s_{23},s_{34},s_{45},s_{15}$)}   & \shortstack{\\Non-adjacent\\invariants\\($s_{13},s_{24},s_{35},s_{14},s_{25}$)}  \\ 
 \hline  \hline
 & & & & \\
 1  & 1,2 & 3,4,5 & $s_{12}, s_{34}, s_{45} >0$  & $s_{35} >0$ 
  \\ 
  &  &  & $s_{23}, s_{51} <0$ &  $s_{13}, s_{15} ,s_{24}, s_{25} <0$  \\ 
 \hline 
  & & & & \\
 2  & 5,1 & 2,3,4 & $s_{51}, s_{23}, s_{34} >0$ & $s_{24} >0$ \\ 
  &  &  & $s_{12}, s_{45} <0$ & $s_{25}, s_{35} ,s_{13}, s_{14} <0$  \\ 
 \hline 
  & & & & \\
 3  & 4,5 & 1,2,3 & $s_{45}, s_{12}, s_{23} >0$ & $s_{13} >0$  \\ 
  &  &  & $s_{51}, s_{34} <0$ & $s_{14}, s_{24} ,s_{25}, s_{35} <0$  \\ 
 \hline 
  & & & & \\
 4  & 3,4 & 5,1,2 & $s_{34}, s_{51}, s_{12} >0$ & $s_{25} >0$  \\ 
  &  &  & $s_{45}, s_{23} <0$   & $s_{35}, s_{13} ,s_{14}, s_{24} <0$  \\ 
 \hline 
  & & & & \\
 5  & 2,3 & 4,5,1 & $s_{23}, s_{45}, s_{51} >0$ & $s_{14} >0$ 
 \\ 
  &  &  & $s_{34}, s_{12} <0$ & $s_{24}, s_{25} ,s_{35}, s_{13} <0$  \\ 
 \hline  \hline \hline
  & & & & \\
 6  & 3,5 & 1,2,4 & $s_{12} > 0$ & $s_{35}, s_{14}, s_{24} >0$  \\ 
  &  &  & $s_{23}, s_{34}, s_{45}, s_{51} < 0$  &  $s_{25}, s_{13} <0$  \\ 
 \hline 
  & & & & \\
 7  & 1,4 & 2,3,5 & $s_{23} > 0$ & $s_{14}, s_{25}, s_{35} >0$  \\ 
  &  &  & $s_{34}, s_{45}, s_{51}, s_{12} < 0$  & $s_{13}, s_{24} <0$  \\ 
 \hline 
  & & & & \\
 8  & 2,5 & 3,4,1 & $s_{34} > 0$ & $s_{25}, s_{13}, s_{14} >0$  \\ 
  &  &  & $s_{45}, s_{51}, s_{12}, s_{23} < 0$   & $s_{24}, s_{35} <0$  \\ 
 \hline 
  & & & & \\
  9  & 1,3 & 4,5,2 & $s_{45} > 0$ & $s_{13}, s_{24}, s_{25} >0$  \\ 
  &  &  & $s_{51}, s_{12}, s_{23}, s_{34} < 0$ &  $s_{14}, s_{35} <0$  \\
 \hline 
  & & & & \\
 10  & 2,4 & 5,1,3 & $s_{51} > 0$ & $s_{24}, s_{35}, s_{13} >0$  \\ 
  &  &  & $s_{12}, s_{23}, s_{34}, s_{45} < 0$   & $s_{25}, s_{14} <0$  
  \\ 
 \hline 
\end{tabular}
\end{center}
\caption{Kinematical channels in Minkowski region: the first five correspond to adjacent channels, while the remaining five 
are non-adjacent channels.}
\label{tab:channels}
\end{table}

Individual five-point integrals may be related among different channels through momentum permutations. 
The kinematical region in each channel is delimited by requiring  all $s$-channel invariants positive and all $t$-channel 
invariants negative, plus negativity $\Delta\leq 0$
 of the Gram determinant (following from the real-valuedness of all momenta). 

\begin{figure}[t!]
\centering
  \includegraphics[width=10cm]{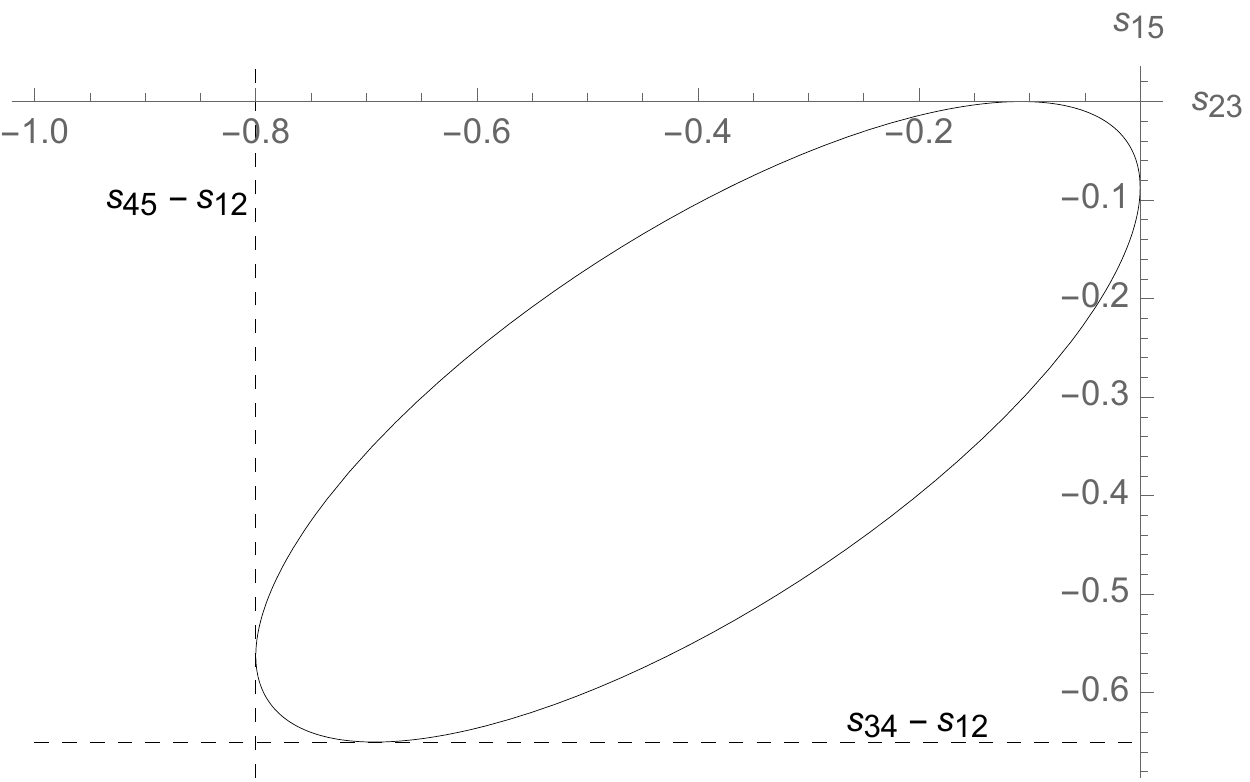}
  \caption{Kinematical region in the $s_{12}$-channel in the $(s_{23},s_{15})$-plane
   for $s_{12}=1.0$, $s_{34}=0.35$, $s_{45}=0.2$ fixed. }
   \label{fig:ellipse}
\end{figure}

On the example of the $s_{12}$-channel, these  non-Gram-determinant constraints imply for the independent invariants:
\begin{equation}
s_{12}\geq s_{34}\,, \qquad s_{12}-s_{34}\geq s_{45}\,, \qquad 0\geq s_{23}\geq s_{45}-s_{12}\,.
\end{equation}
The last remaining independent invariant is then constrained by the positivity of the Gram determinant:
\begin{equation}
s_{15}^+ \geq s_{15} \geq s_{15}^-\,,
\end{equation}
with 
\begin{eqnarray}
s_{15}^\pm &=& \frac{1}{(s_{12}-s_{45})^2} \bigg( s_{12}^2s_{23} + s_{34}s_{45}(s_{45}-s_{23}) - s_{12}(s_{34}s_{45}
+ s_{23}s_{34} + s_{23}s_{45}) \nonumber \\
&& \hspace{6mm} \pm \sqrt{s_{12}s_{23}s_{34}s_{45}(s_{12}+s_{23}-s_{45})(s_{34}+s_{45}-s_{12})} \bigg).
\end{eqnarray}
For fixed values of $s_{12},s_{34},s_{45}$, these constraints describe an ellipse in the $(s_{23},s_{15})$-plane, as 
shown in an example in Figure~\ref{fig:ellipse}.
\clearpage

\section{Two-loop five-point planar master integrals}
\label{sec:mi}
\begin{figure}[t]%
\center{\includegraphics[width=0.50\linewidth]{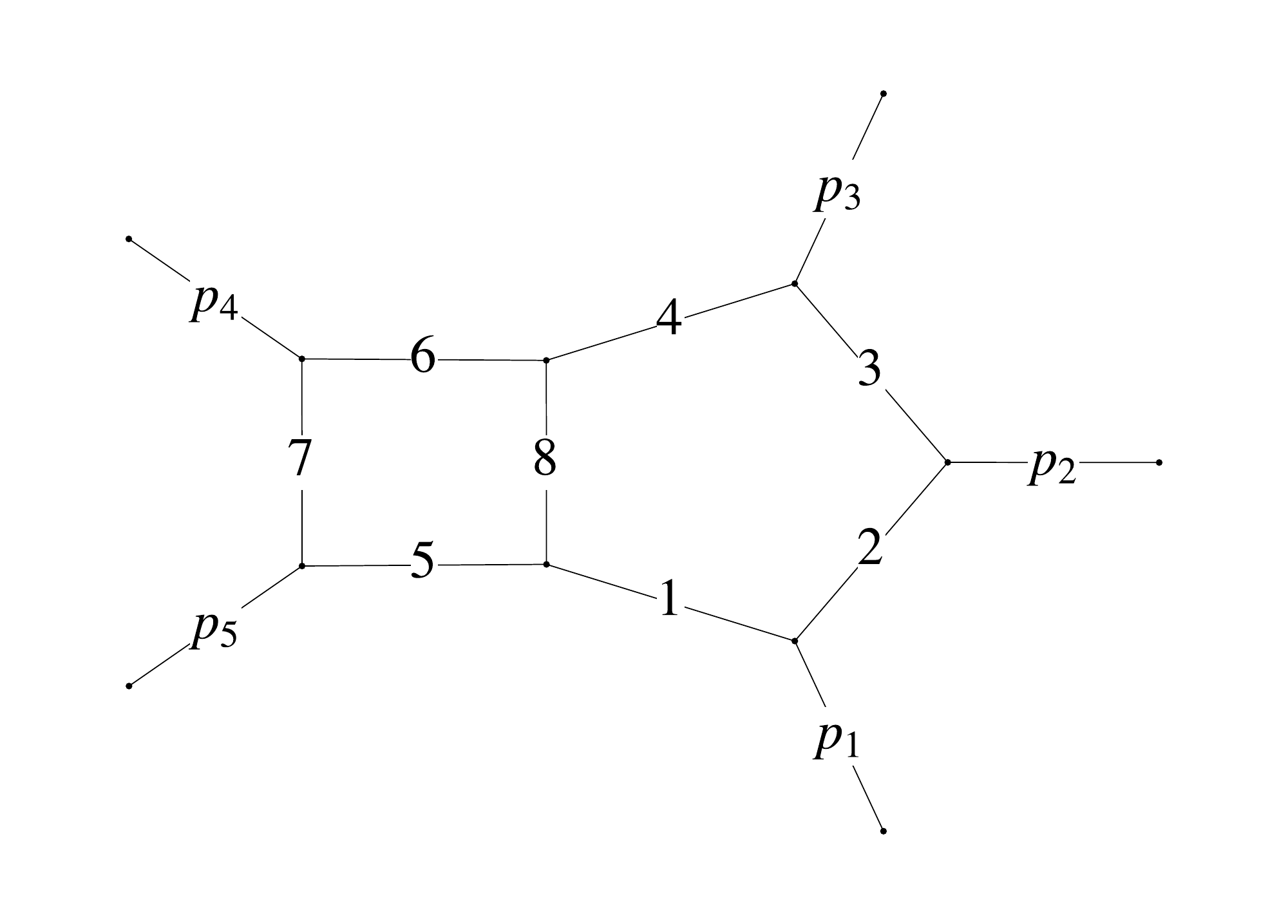}}
  \caption{Family of planar penta-box integrals computed in the main text. The numbers correspond to indices $i$ of $a_{i}$ in $G_{a_{1} \ldots a_{11}}$. Numerator factors are not shown in the figure.}
  \label{figpb}
\end{figure}

The family of penta-box integrals is defined as
\begin{align}
G_{a_1, \ldots a_{11}} :=& \int \frac{d^{D}k_{1} d^{D}k_{2}}{(i \pi^{D/2})^2}\times \nonumber \\&
\times  \frac{[-(k_{1}+p_{1}+p_{2}+p_{3}+p_{4})^2]^{-a_{9}}}{[-k_{1}^2]^{a_1} [-(k_{1}+p_{1})^2]^{a_2}  
[-(k_{1}+p_{1}+p_{2})^2]^{a_3}[-(k_{1}+p_{1}+p_{2}+p_{3})^2]^{a_4}} \times \nonumber \\
&  \times   \frac{[-(k_{2}+p_{1})^2]^{-a_{10}}  }{[-k_{2}^2]^{a_5}   [-(k_{2}+p_{1}+p_{2} +p_{3})^2]^{a_6}[-(k_{2}+p_{1}+p_{2}+p_{3}+p_{4})^2]^{a_7}} 
\times \nonumber \\ \times &
\frac{ [-(k_{2}+p_{1}+p_{2})^2]^{-a_{11}}}{[-(k_1-k_2)^2]^{a_{8}}}\,,
\end{align}
with $p_{i}^2=0, i=1,\ldots 5$, and $\sum_{i=1}^{5} p^\mu_{i} =0$, and where $a_{1} ,\ldots a_{8} \ge 0$ are propagators and $a_{9},a_{10},a_{11} \le 0$ numerator factors.
See Figure \ref{figpb}.

\begin{figure}[t]
 \centerline{
   \subfloat[$I_{6},I_{11}$]{\includegraphics[scale=0.25]{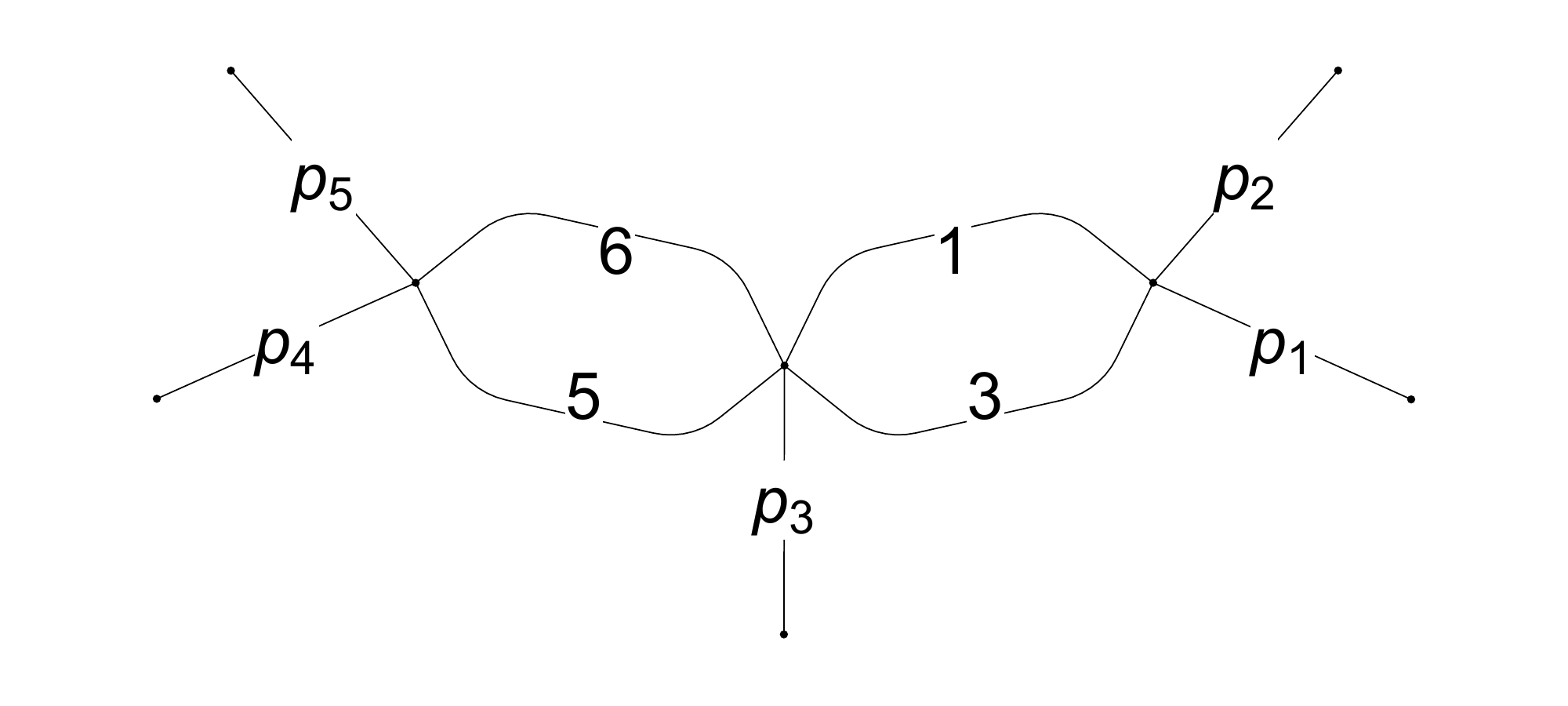}}
    \subfloat[$I_{9}$]{\includegraphics[scale=0.25]{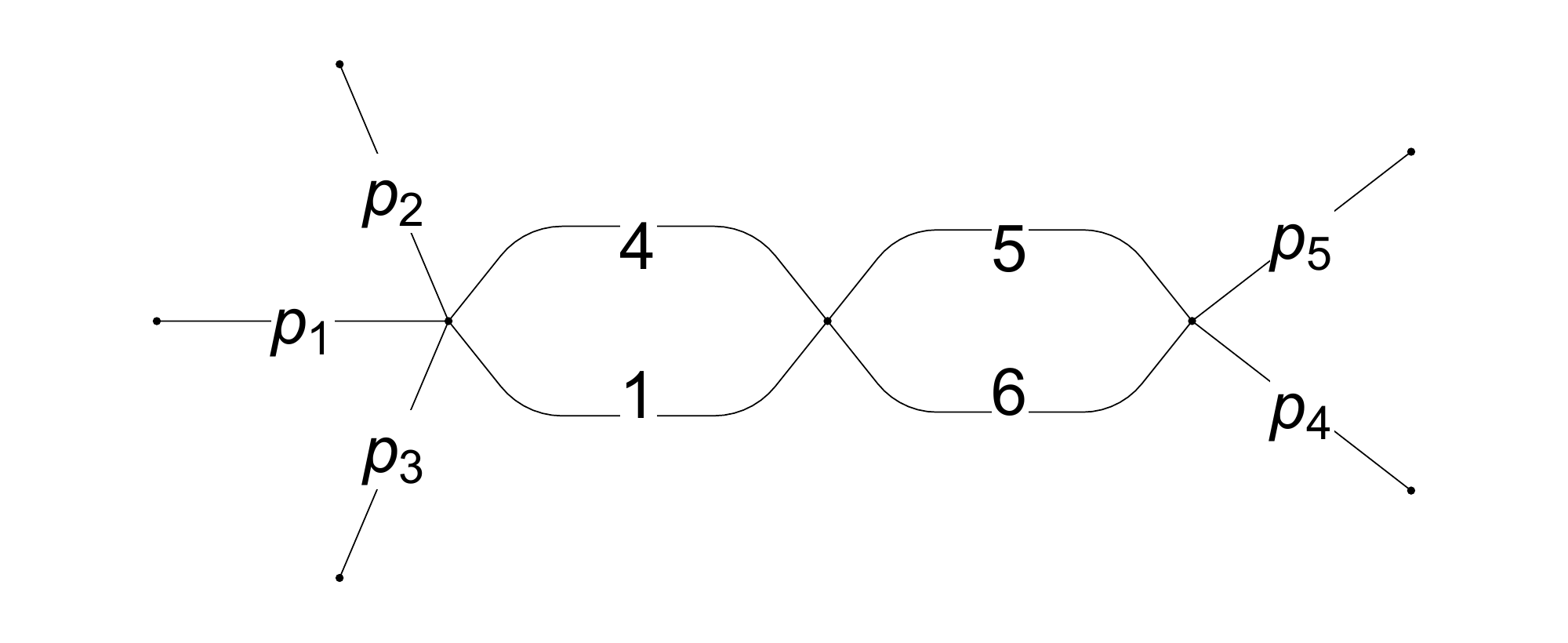}}
       \subfloat[$I_{36}$]{\includegraphics[scale=0.25]{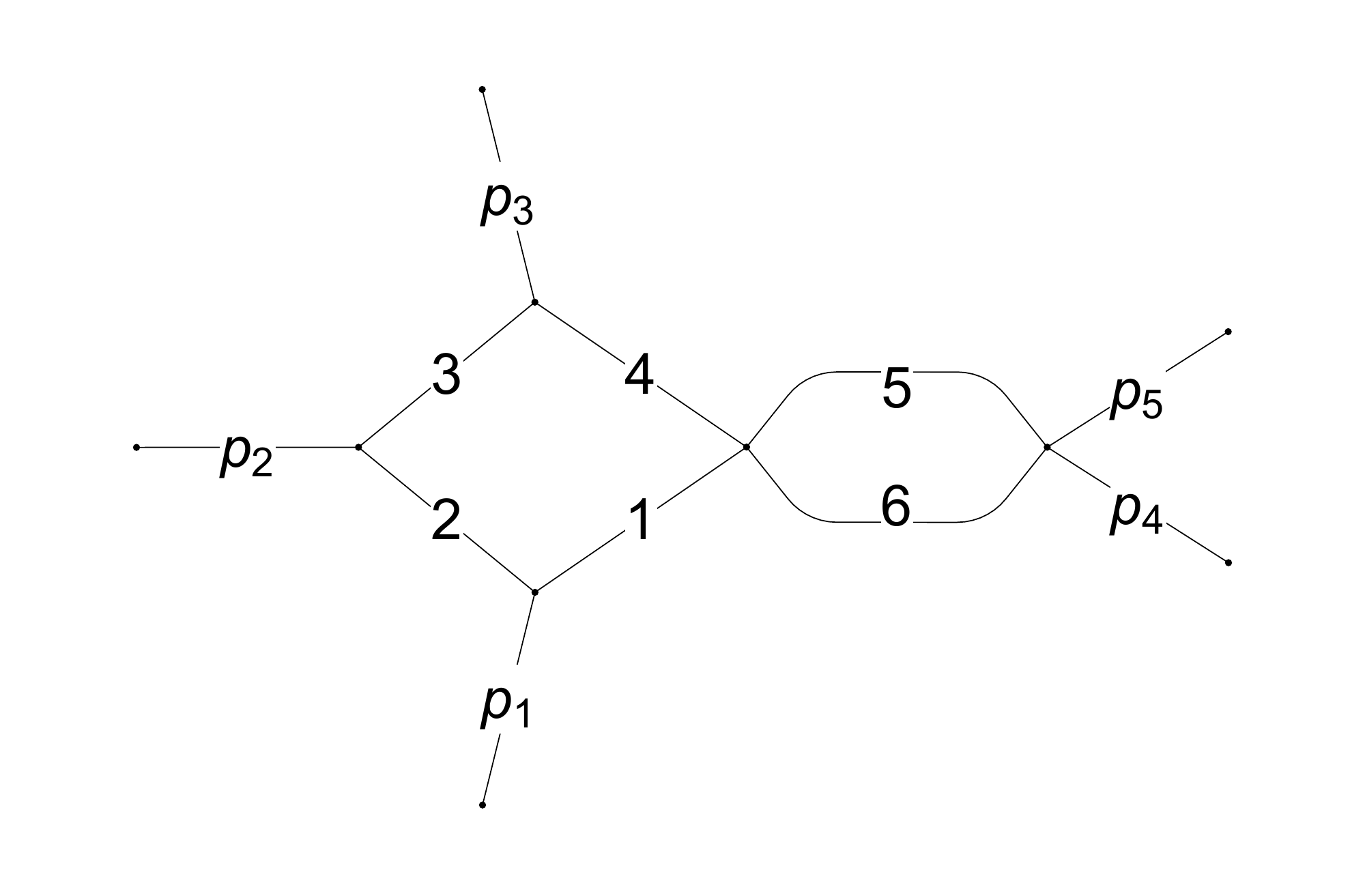}}
 }
   \caption{Factorized integrals.}
 \label{fig:allpentagonintegralsfactorized}
\end{figure}

\begin{figure}[t]
 \centerline{
  %\hspace{-0.3cm}
  \subfloat[$I_{1},I_{2},I_{3},I_{4},I_{5}$]{\includegraphics[scale=0.25]{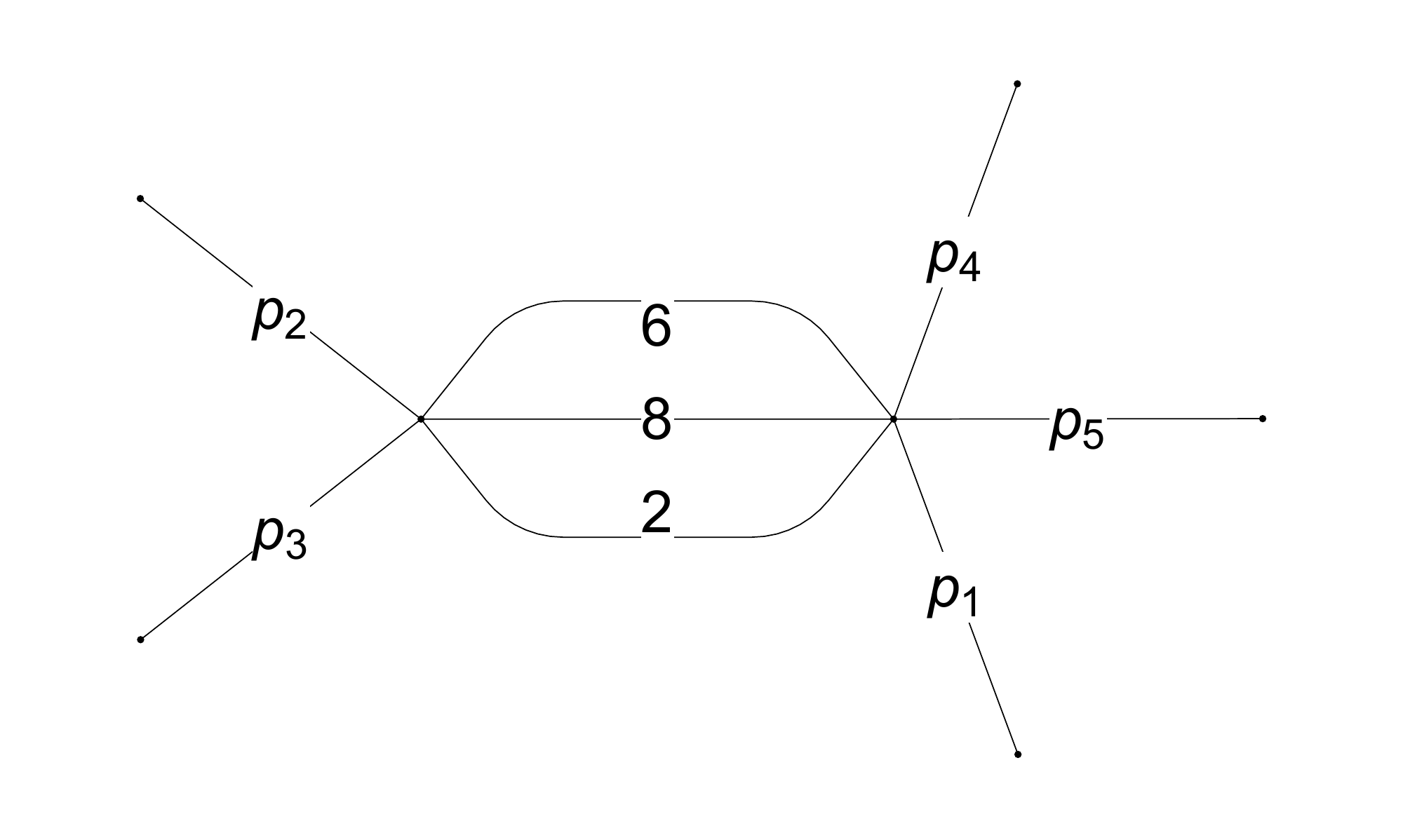}}
      \subfloat[$I_{7},I_{8},I_{12},I_{13},I_{14},I_{15}$]{\includegraphics[scale=0.25]{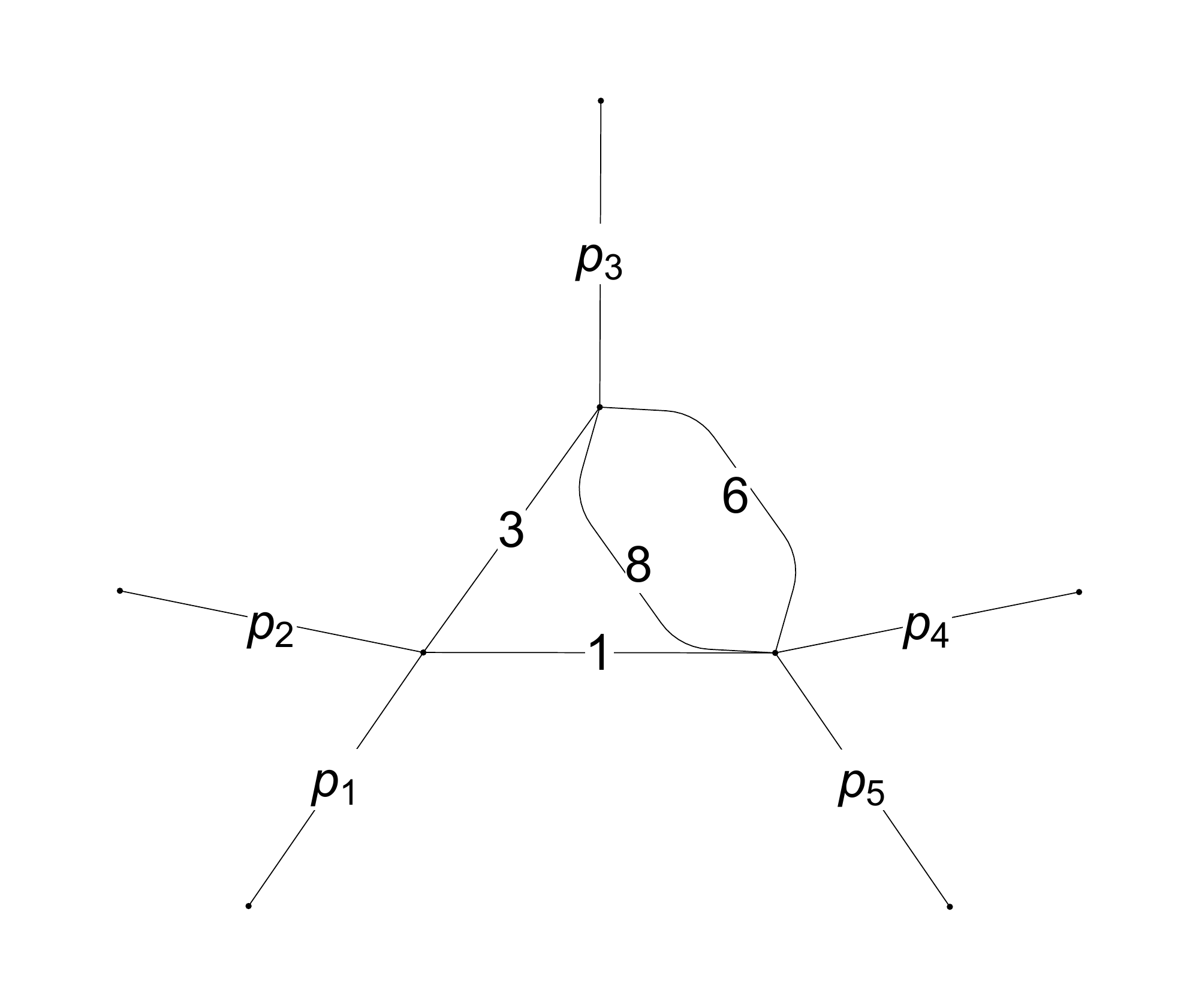}}
       \subfloat[$I_{10}$]{\includegraphics[scale=0.25]{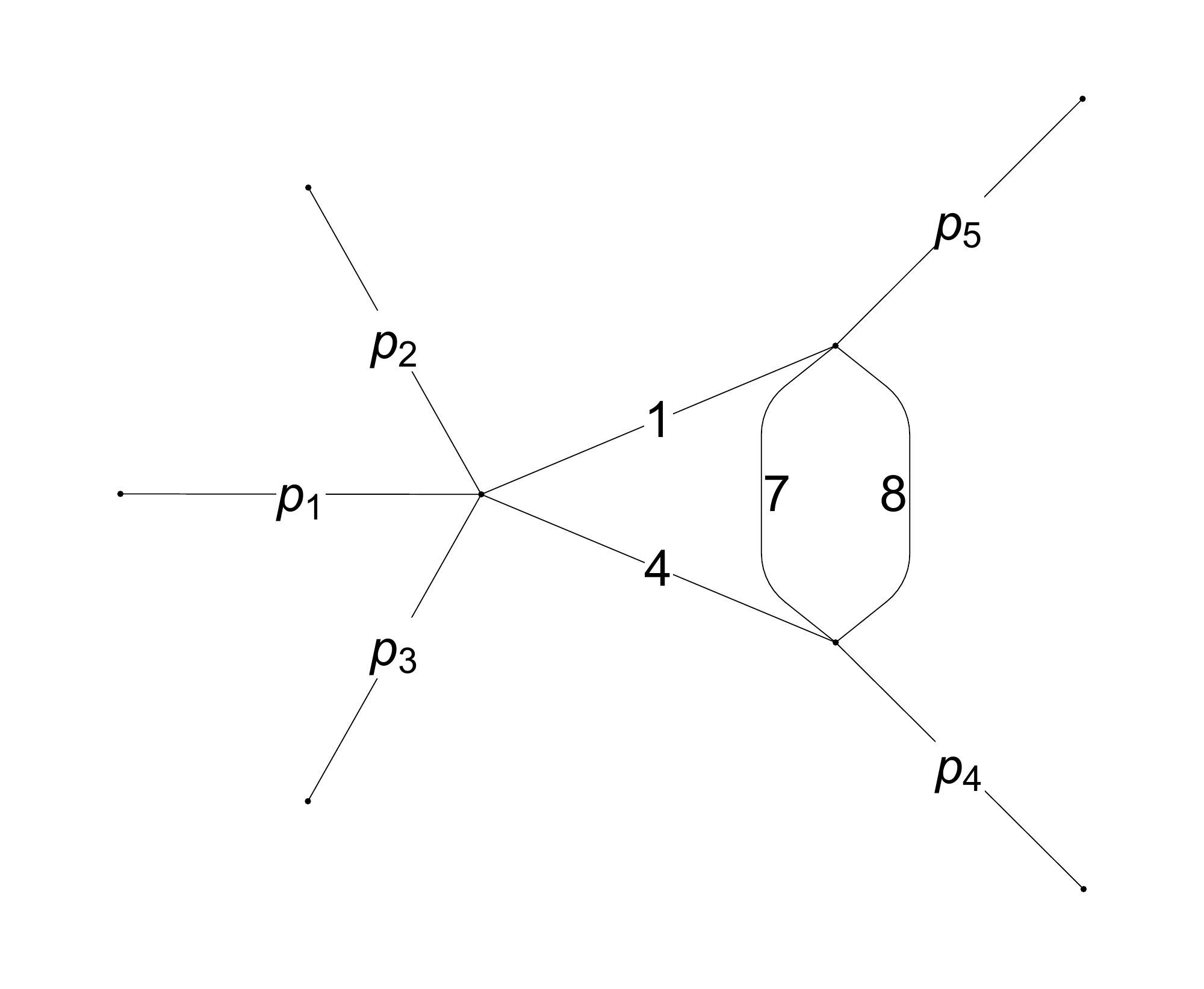}}
 }
%  \vspace{-0.2cm}
 \centerline{
     \subfloat[$I_{16},I_{17},I_{24},I_{25},I_{33},I_{35}$]{\includegraphics[scale=0.25]{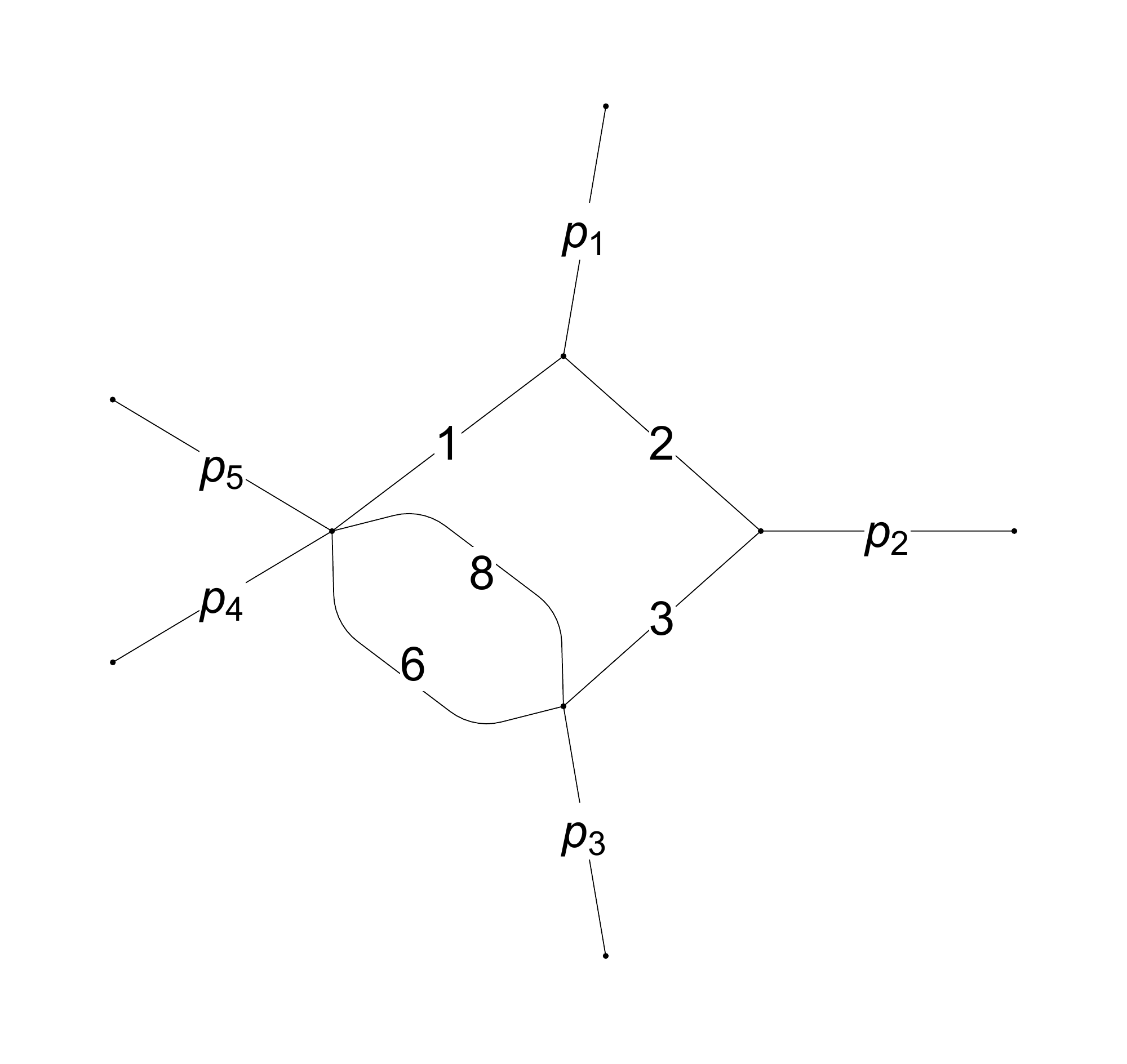}}
       \subfloat[$I_{18},I_{20}$]{\includegraphics[scale=0.25]{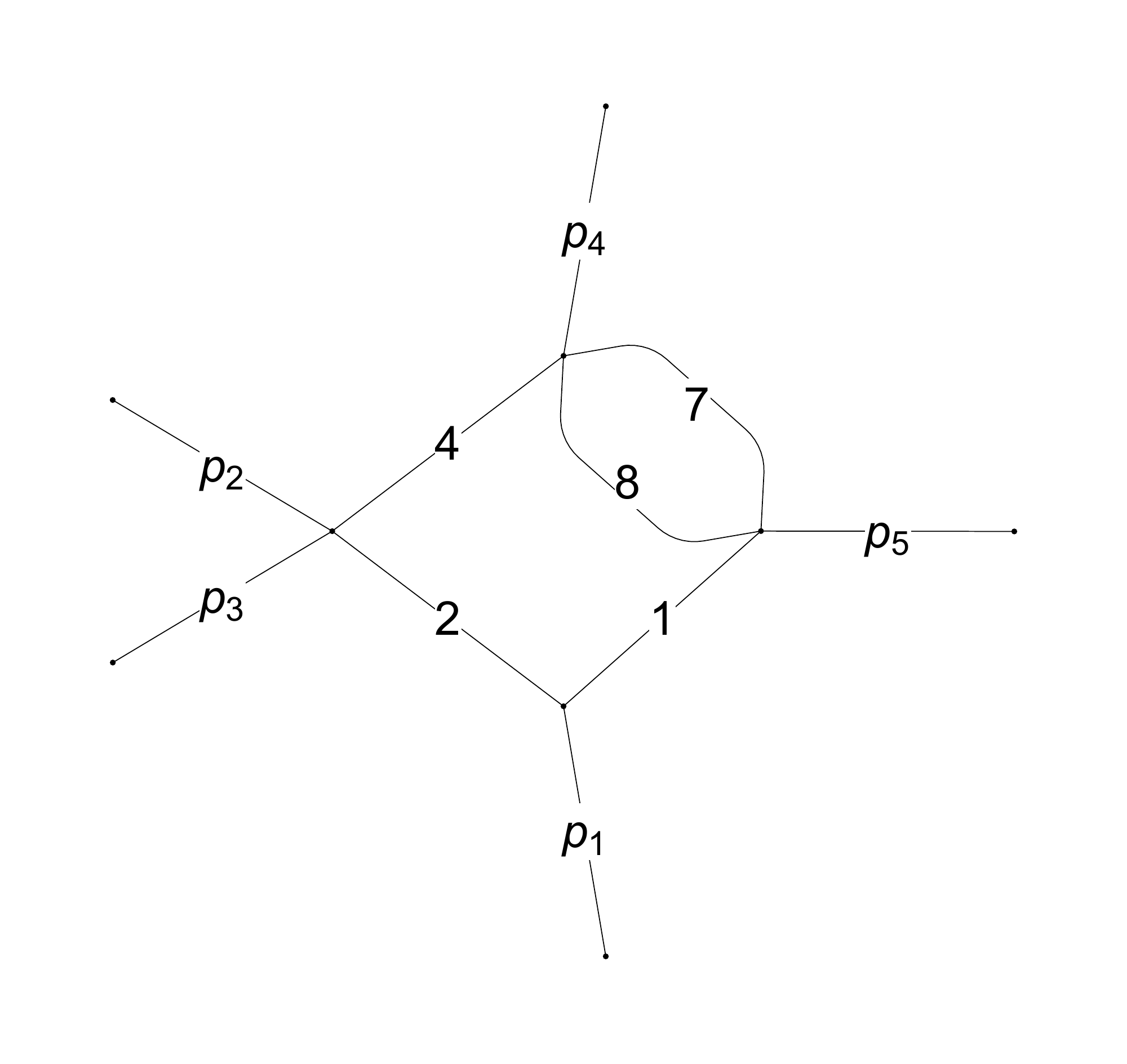}}
       \subfloat[$I_{19},I_{28},I_{29},I_{34}$]{\includegraphics[scale=0.25]{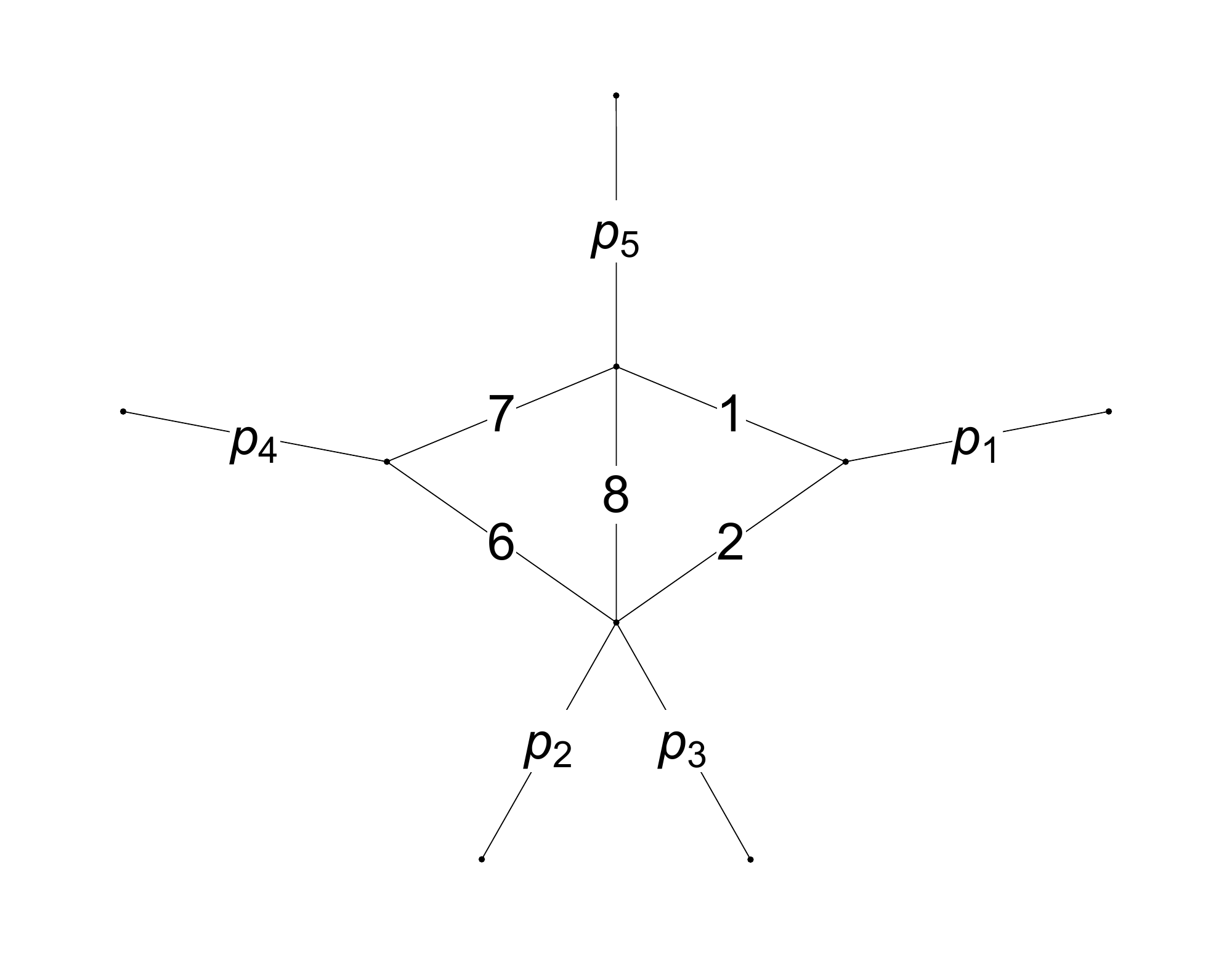}}
         }  \centerline{
       \subfloat[$I_{21},I_{30}$]{\includegraphics[scale=0.25]{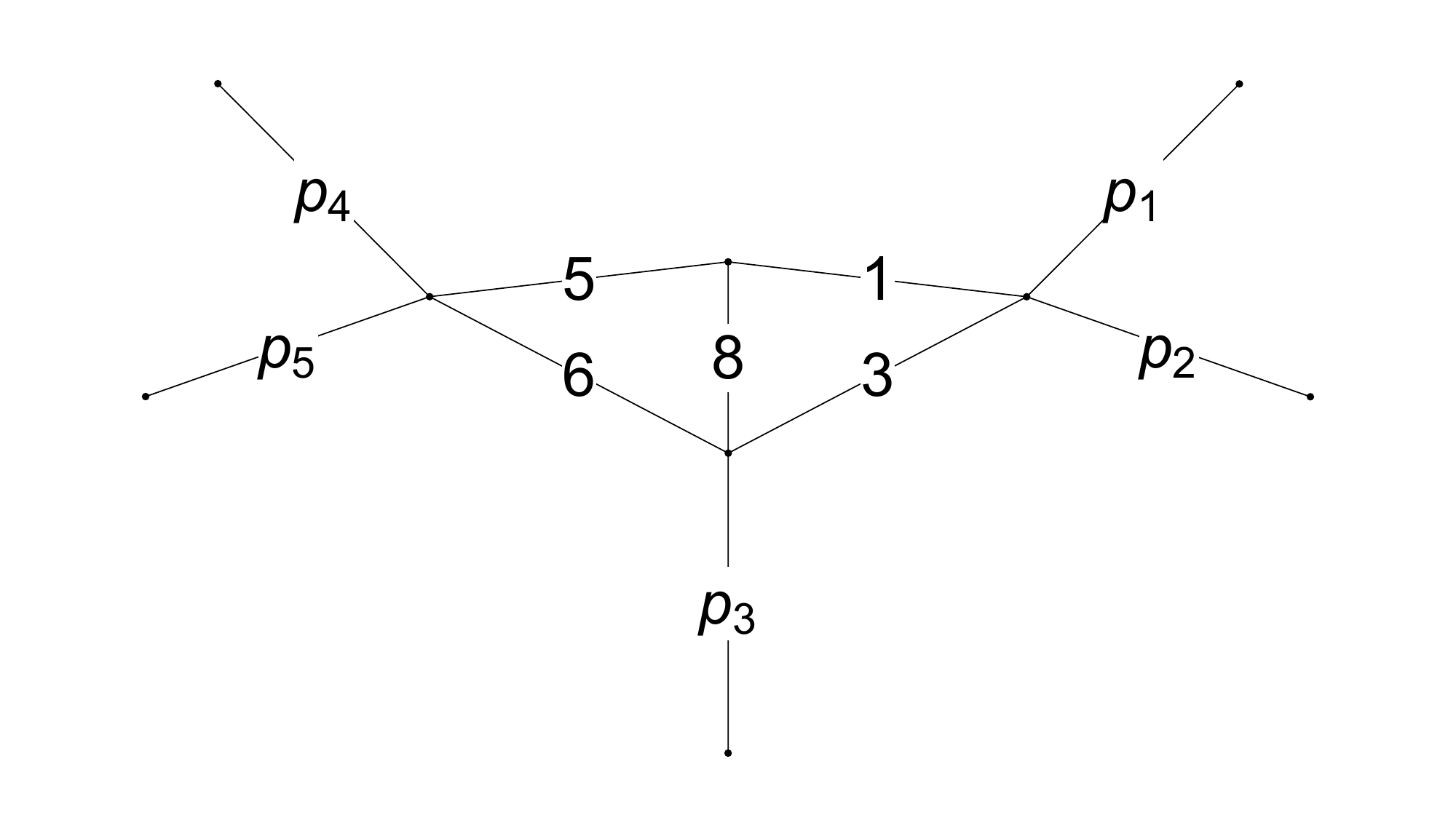}}
       \subfloat[$I_{22},I_{23},I_{26},I_{27},I_{31},I_{32}$]{\includegraphics[scale=0.25]{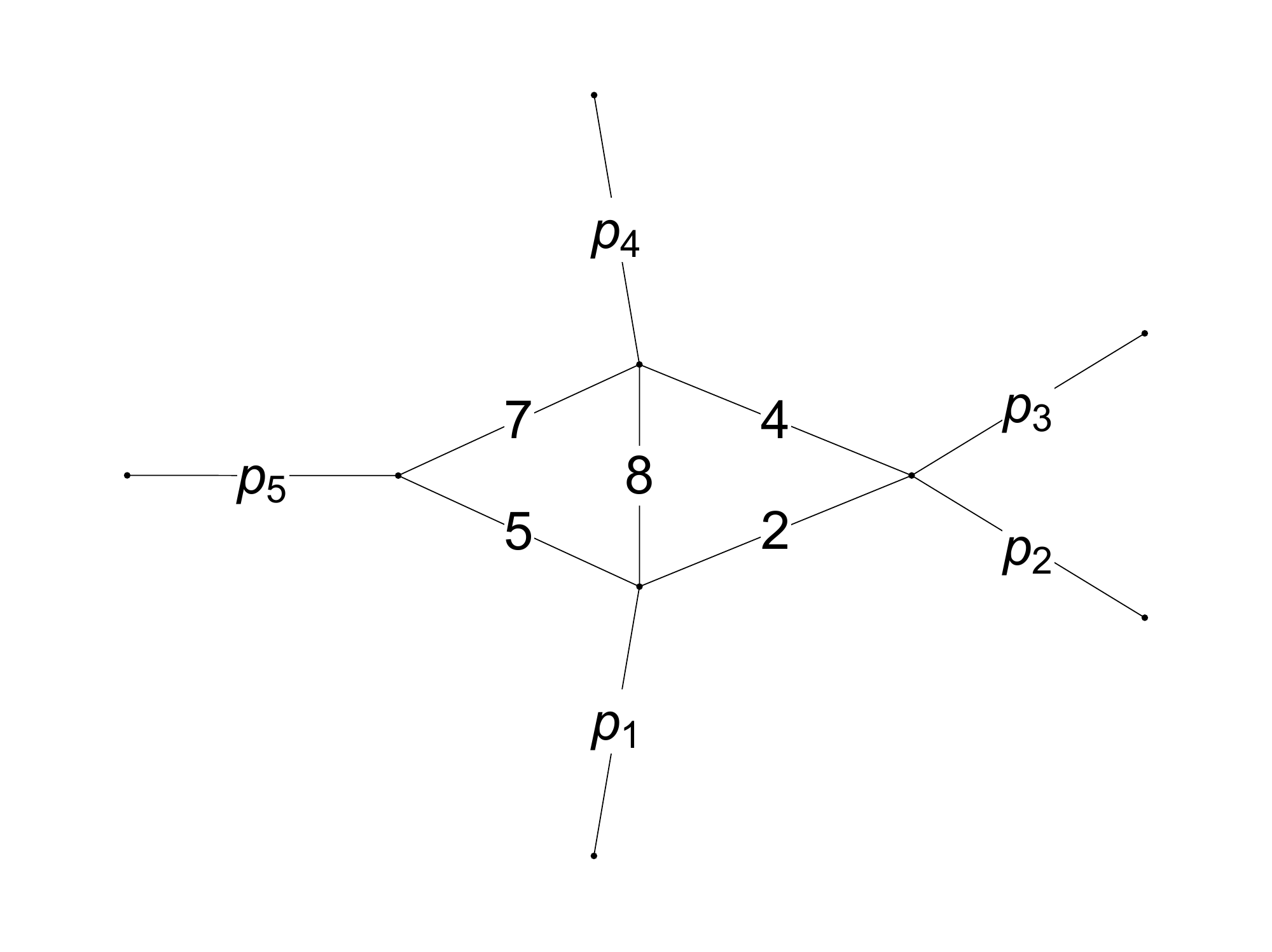}}
       \subfloat[$I_{39},I_{42},I_{43},I_{48}$]{\includegraphics[scale=0.25]{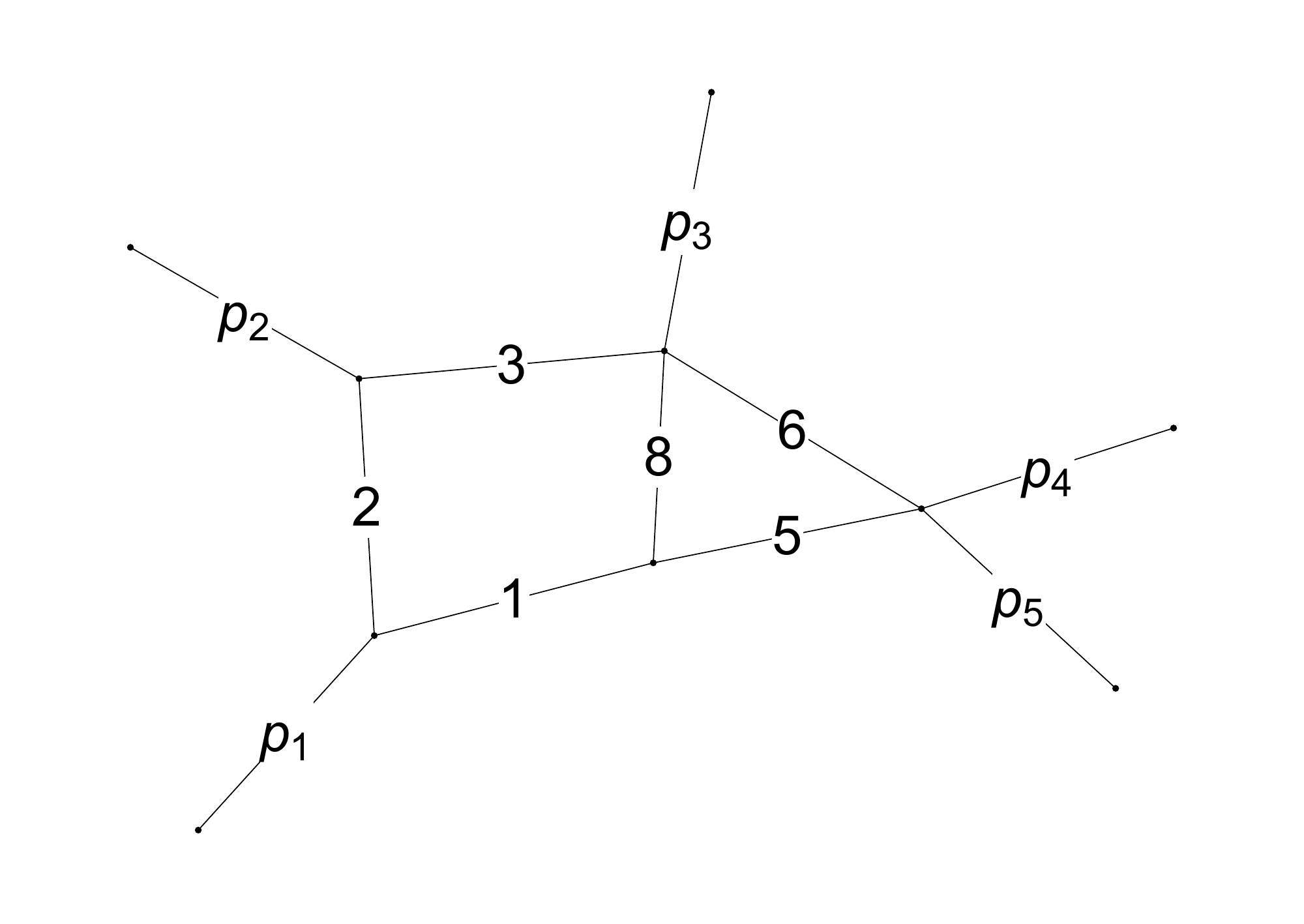}}
         }  \centerline{
       \subfloat[$I_{52},I_{53},I_{54},I_{55}$]{\includegraphics[scale=0.25]{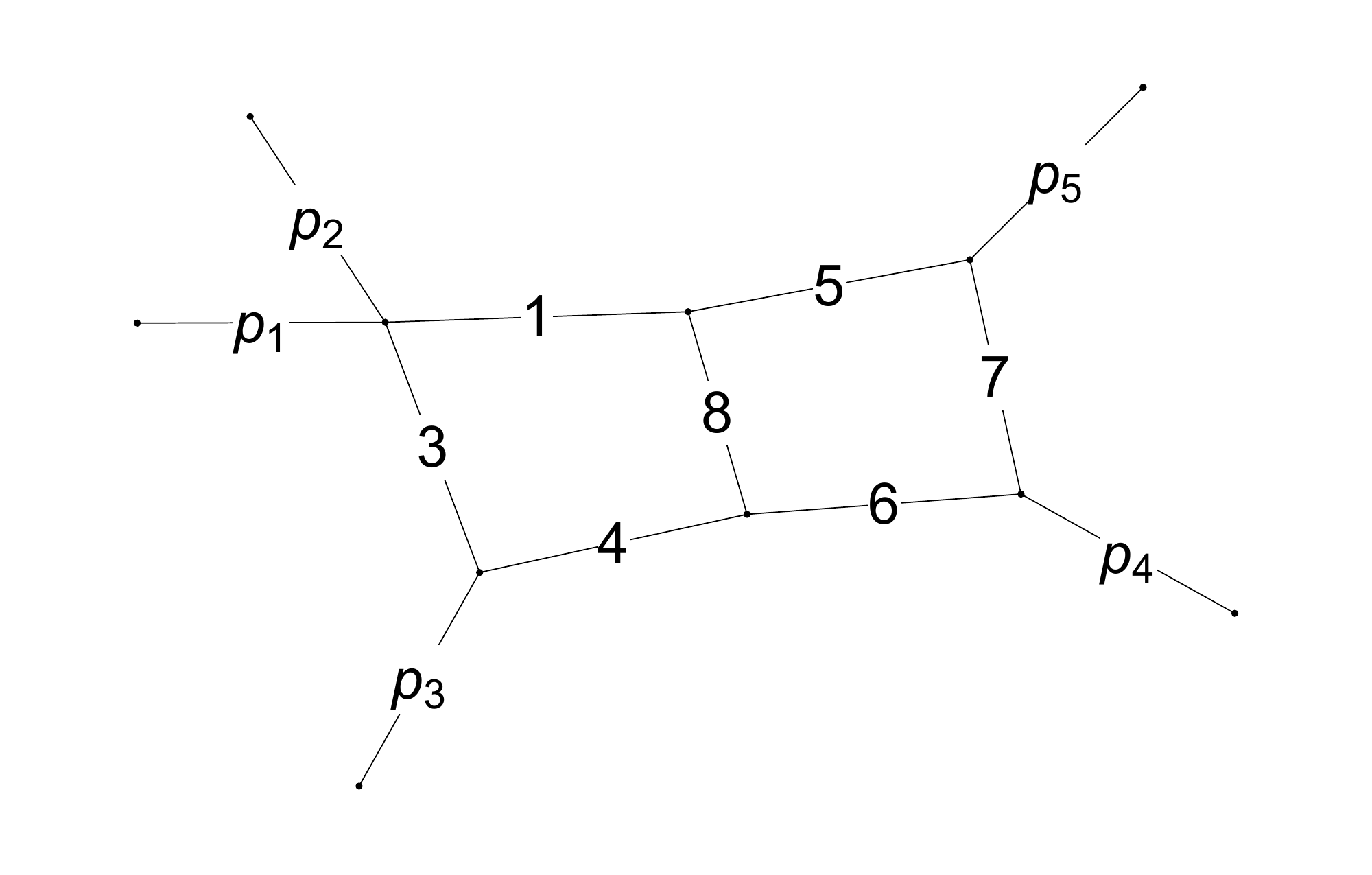}}
 }
  \caption{Five-particle integrals in four-point kinematics. Graphs related by symmetries are not shown.}
 \label{fig:allpentagonintegrals4pt}
\end{figure}

\begin{figure}[t]
 \centerline{
  \subfloat[$I_{37},I_{38}$]{\includegraphics[scale=0.20]{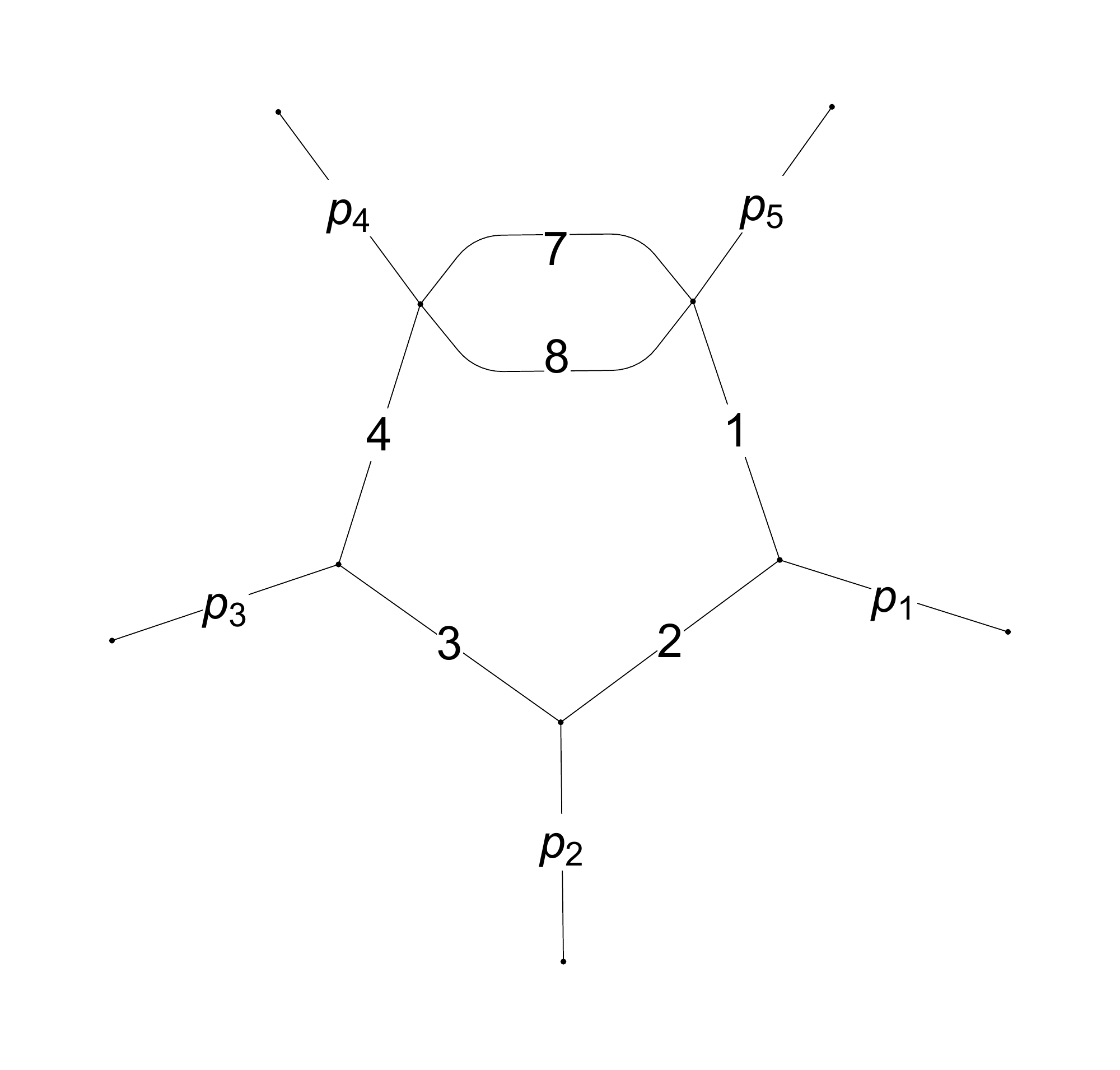}}
  \subfloat[$I_{40},I_{41},I_{44},I_{45},I_{46},I_{47}$]{\includegraphics[scale=0.20]{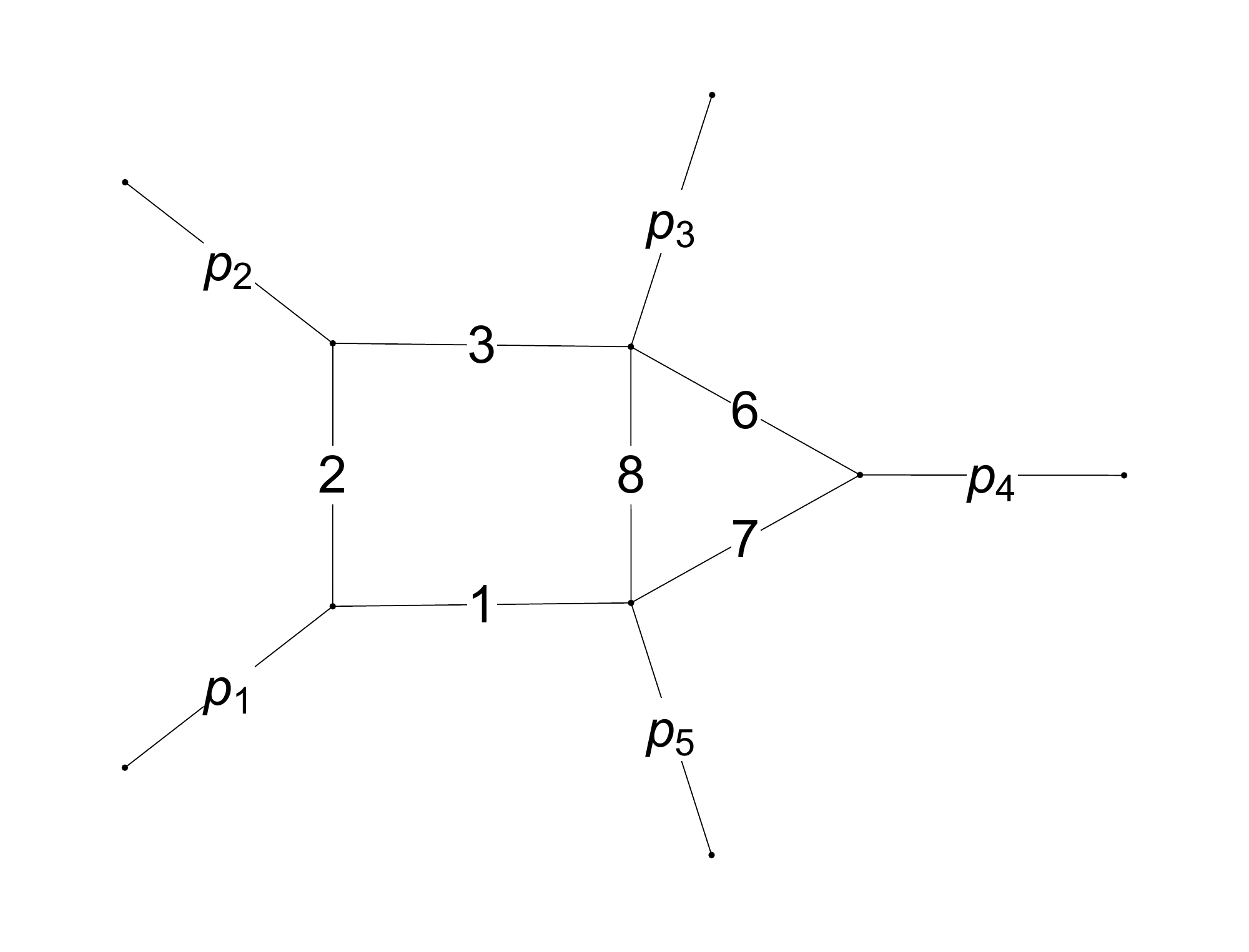}}
  \subfloat[$I_{49},I_{50},I_{51},I_{56},I_{57},I_{58}$]{\includegraphics[scale=0.20]{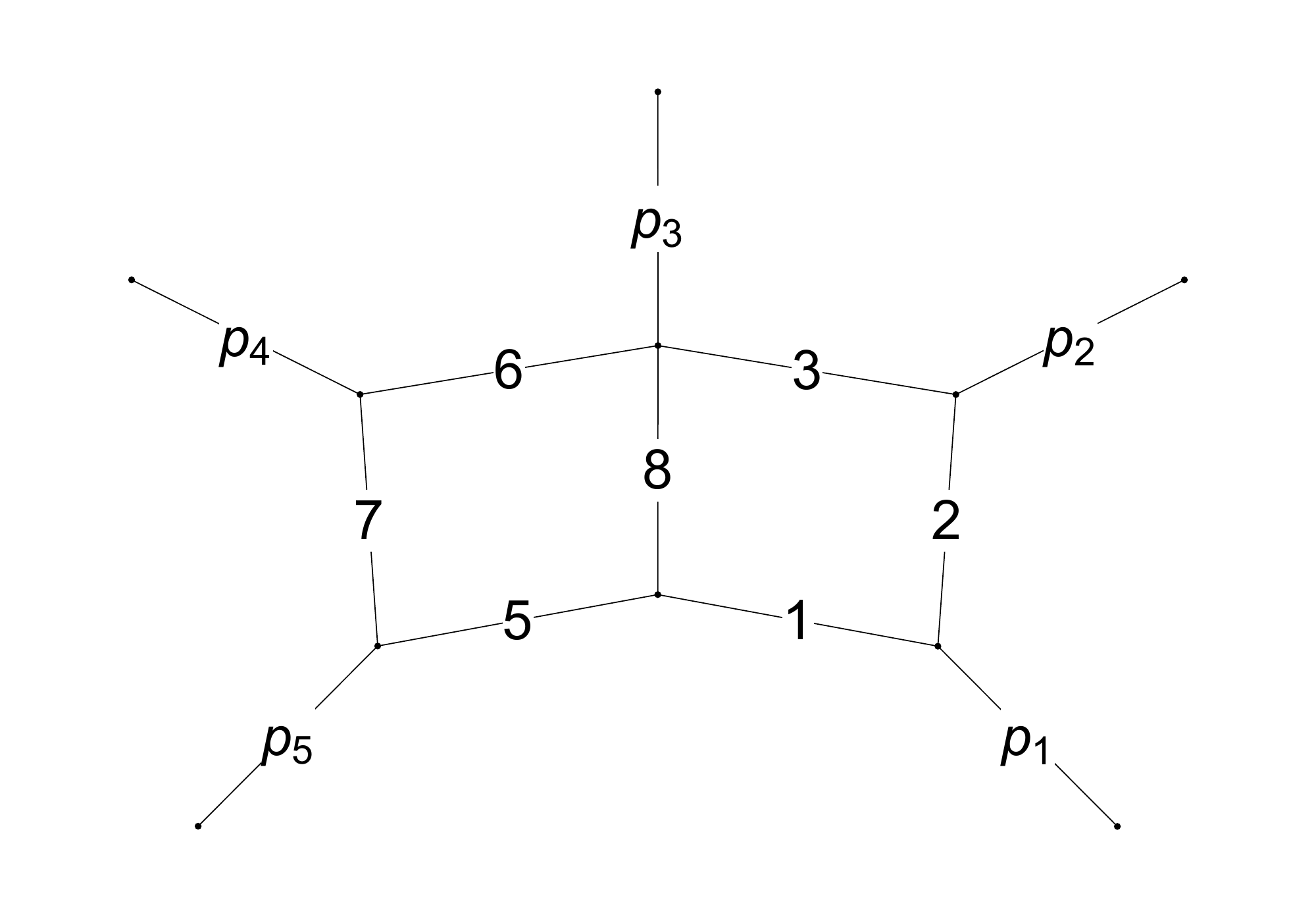}}
       \subfloat[$I_{59},I_{60},I_{61}$]{\includegraphics[scale=0.20]{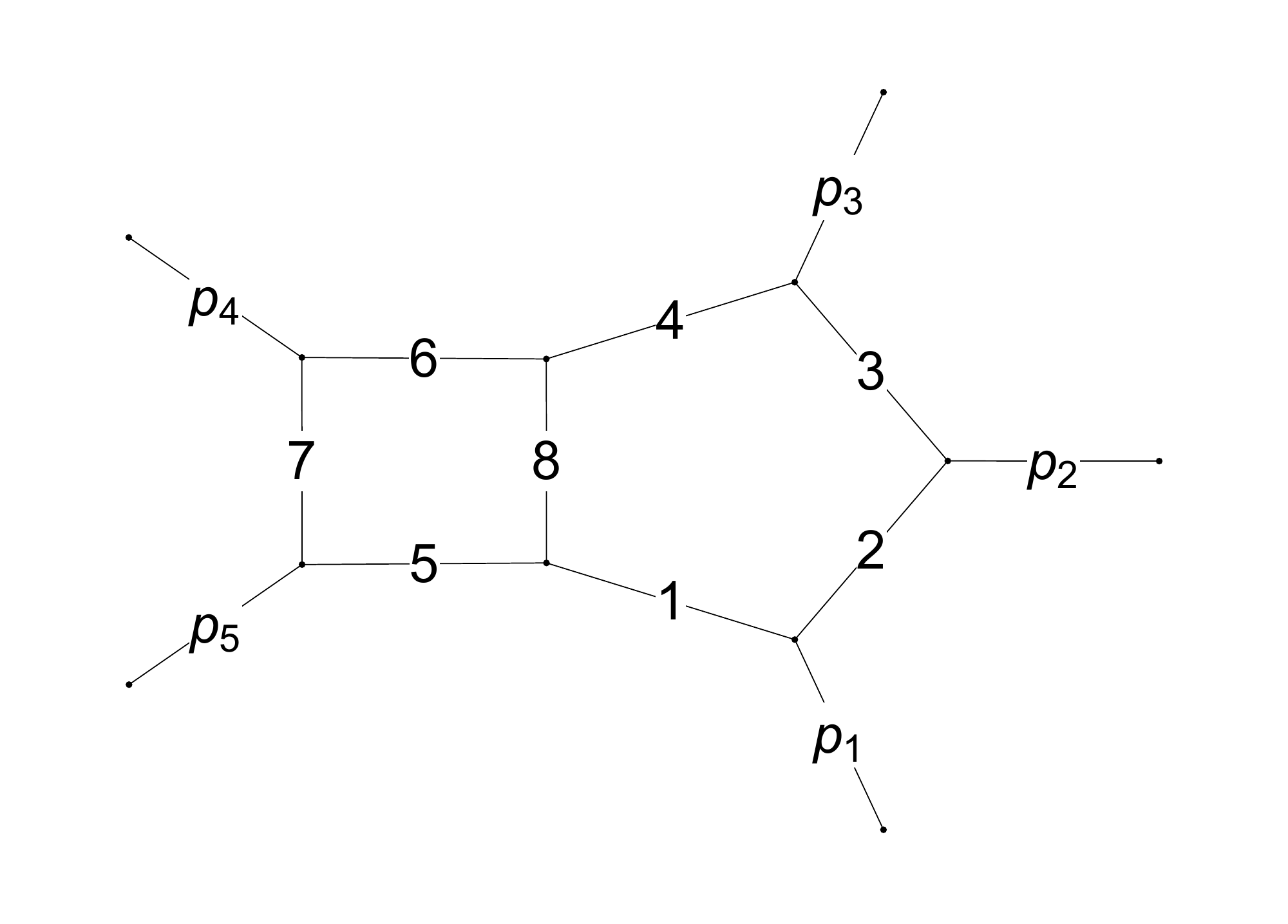}}
 }
   \caption{Genuine five-particle integrals.}
 \label{fig:genuine5pt}
\end{figure}

Integral reduction~\cite{Laporta:2001dd} (using for example FIRE~\cite{Smirnov:2008iw} or 
Reduze~\cite{vonManteuffel:2012np})
 shows that there are $61$ master integrals for this family of integrals.
The master integrals can be organized in terms of integral sectors, corresponding to the $8$-propagator sector,
and subsectors with fewer propagators.  In total, one needs $46$ distinct sectors. One can further organize the latter
by grouping together sectors that are related by relabelling the external legs. In this way, one can group all integrals
in terms of $17$ sectors, which are classified in the following. 
First, there are a number of integrals that are products of one-loop integrals. They are shown in Fig.~\ref{fig:allpentagonintegralsfactorized}.
In this figure, the $I_{j}$ indicate at which position these integrals appear in the basis that we chose (the latter will be discussed below).
Next, integrals corresponding to four-point functions with one off-shell leg are shown in Fig.~\ref{fig:allpentagonintegrals4pt}.
They are known from ref.~\cite{Gehrmann:2000zt,Gehrmann:2002zr}, where they are expressed in terms of generalized 
harmonic polylogarithms~\cite{Remiddi:1999ew,2001math3059G}, for which efficient numerical evaluations 
are available~\cite{Gehrmann:2001pz,Gehrmann:2001jv,Vollinga:2004sn}. 
The genuine five-point sectors are shown in Fig.~\ref{fig:genuine5pt}.

For what follows, we need to choose a basis of $61$ integrals. Traditionally, integrals were typically chosen relatively randomly, as implied by the lexicographic ordering~\cite{Laporta:2001dd} that was used in the integral reduction. 
In ref.~\cite{Henn:2013pwa}, it was proposed to choose the basis such that the integrals have simple properties.
This considerably simplifies their computation, and allows to obtain the result in a form that is as compact as possible.
As we will see in the next section, in the $\eps$ expansion, all integrals evaluate to multiple polylogarithms.
In general, such Feynman integrals will involve linear combinations of multiple polylogarithms of varying transcendental weight, and with various prefactors that depend rationally or algebraically on the kinematics.
It is desirable to disentangle the latter, so that such factors are moved into overall normalizations of the integrals, and such that only functions of homogeneous weight appear.

Understanding for predicting which integrals have this property came initially from studies in $\mathcal{N}=4$ super Yang-Mills.
Conjecturally, integrals whose  {\it integrands} can be written as a `d-log'-form, and hence have constant leading singularities  \cite{ArkaniHamed:2010gh},
have this property.  The initial examples satisfying this conjecture were massless, planar,  finite, dual conformal integrals. It has since been generalized to more generic integrals within dimensional regularization \cite{Henn:2013pwa}.
 
What is important to emphasize is that the basis choice can be done {\it a priori}, by analyzing the loop {\it{integrand}}. 
In principle, one could classify all integrands having the desired properties, and then select a linearly independent (under integral reduction) subset.
This can be done algorithmically, see e.g.~\cite{WasserMSc}.

In practice, it may not be necessary to classify all such integrals, but just to construct 
a sufficient number of them. It is possible to construct many `d-log' integrals
directly, for example by iteratively using lower-loop building blocks.
See refs. \cite{Henn:2013fah,Henn:2014qga} for examples.
Our choice of $61$ basis integrals is given in the ancillary file {\tt{pentabox$\_$basis2.txt}}.

\section{Differential equations for the master integrals}
\label{sec:differentialequations}

To compute the master integrals, the explicit  and complicated integration over the loop momenta 
can often be avoided for multi-scale integrals by using  differential equations in 
kinematical invariants, as first demonstrated for the two-loop four-point functions in~\cite{Gehrmann:1999as}.  
 
We use the integral basis $I_{j}$, with $j=1\ldots 61$, discussed in the previous section,
and compute the differential in all variables $v_{j}$, $j=1,\ldots 5$.
 We find  the following canonical form of the differential equations \cite{Henn:2013pwa}
\begin{align}\label{canonicalDEpentagon}
d \vec{I}(v_{i};\epsilon) = \epsilon \, d \tilde{A} \, \vec{I}(v_{i};\epsilon) \,,
\end{align} 
with the matrix
\begin{align}\label{definition-Atilde}
\tilde{A}=  \left[ \sum_{i=1}^{31} a_{i} \,  \log W_{i}(v_{i}) \right] \,.
\end{align}
 The ensemble of the letters $W_i$ is called the alphabet of the 
problem under consideration. We derived the  planar pentagon alphabet in our previous work~\cite{Gehrmann:2015bfy}, 
which was subsequently extended to account also for letters relevant to non-planar pentagon functions 
in ref.~\cite{Chicherin:2017dob}, whose notation we adopt here: 
\begin{align}\label{fullpentagonalphabet}
W_{1} =& \; v_{1} \,,  \\
W_{6} =& \; v_{3} + v_{4} \,,  \ \\
W_{11} =& \; v_{1} - v_{4} \,,  \\ 
W_{16} =& \; v_{4} - v_{1}- v_{2} \,,  \,, \\
W_{21} =& \; v_3 + v_4 - v_1 - v_2 \,,  \\
W_{26} =& \; \frac{v_{1} v_{2} -v_{2} v_{3} +v_{3} v_{4} -v_{1} v_{5} -v_{4} v_{5} - \sqrt{\Delta} }{v_1 v_2 -v_2 v_3 +v_3 v_4 -v_1 v_5 -v_4 v_5 + \sqrt{\Delta}} \,,  \\
W_{31} =& \; \sqrt{\Delta} \,,
\end{align}
with $W_{1+i}, W_{6+i}, W_{11+i}, W_{16+i}, W_{21+i}, W_{26+i}$, with $i=1\ldots 5$, defined by cyclic symmetry.
Note that the $W_{i}$, with $i=26, \ldots 30$, are parity-odd, in the sense that they go to their inverse under $\Delta \to - \Delta$,
while all other letters are parity-even under that transformation.
Finally, the $a_{i}$ are {\it constant} $61 \times 61$ matrices.

The $W_{i}$ notation covers that case of planar \cite{Gehrmann:2015bfy} as well as non-planar pentagon functions \cite{Chicherin:2017dob}.
Following the notation of that reference, the full $31$-letter alphabet is called $\mathbb{A}_{\rm NP}$.
Here we will only need the planar case, which consists of the $26$ letters $\{ W_{1}\,, \ldots W_{20} \,, W_{26} \,,\ldots W_{31} \}$, denoted by $\mathbb{A}_{\rm P}$.
We remark that in the one-loop case, only $21$ letters are needed, namely $\{ W_{1}\,,\ldots  W_{5}\,, W_{11} \ldots W_{20} \,, W_{26} \,,\ldots W_{31} \}$. We denote this alphabet by $\mathbb{A}_{{\rm P};1}$.

The planar $26$-letter alphabet $\mathbb{A}_{\rm P}$ covers the pentabox integrals (and all their subintegrals), as well as cyclic rotations thereof. 
If one focuses only on the orientation of the pentabox shown in Fig.~\ref{figpb}, only $24$ of those letters occur; the letters that are absent in that case are $W_{8}$ and $W_{10}$.

\subsection{Chen iterated integrals for solving the differential equation}

Given the first-order system of differential equations (\ref{canonicalDEpentagon}), the solution is completely determined upon giving a boundary condition. It turns out that the latter can be obtained by simple physical considerations. This will be discussed in section \ref{sec:boundary}.

What functions will appear in the solution of eq. (\ref{canonicalDEpentagon})? In practice, we need to solve this equation in an expansion in $\epsilon$, i.e.
\begin{align}\label{UT-integrals-eps-expansion}
\vec{I}(v_{i};\epsilon)  = \sum_{k \ge 0} \eps^k \vec{I}^{(k)}(v_{i}) \,.
\end{align}
Note that our integral basis $\vec{I}$ is normalized, without loss of generality, such that this expansion starts at $\eps^0$.

Inserting (\ref{UT-integrals-eps-expansion}) into (\ref{canonicalDEpentagon}), we see that the equation decouples order by order in $\eps$,
\begin{align}
d \vec{I}^{(k+1)}(v_{i}) = (d \tilde{A}(v_{i}) ) \vec{I}^{(k)}(v_{i}) \,.
\end{align}
This means that we have
\begin{align}\label{DE-iterative-solution}
\vec{I}^{(k+1)}(v_{i}) = \int_{\mathcal{C}}   (d \tilde{A}(v'_{i}) ) \vec{I}^{(k)}(v'_{i})  + \vec{I}^{(k)}_b \,,
\end{align}
where $\mathcal{C}$ is an integration contour in the space of kinematic variables $v_i$, and $\vec{I}^{(k)}_b$ represents the boundary value.  

We see that the solution, to all orders in the $\eps$ expansion, is given in terms of iterated integrals. The integration kernels are given by logarithmic differential forms, see (\ref{definition-Atilde}). The possible arguments of the logarithms (the letters) are given by the set $\mathbb{A}_{\rm P}$. We call the 
functions arising from this alphabet planar {\it pentagon functions}.
Along the lines of this terminology, let us observe that the matrix $\tilde{A}$ of eq. (\ref{definition-Atilde}) dictates, via 
eq.~(\ref{DE-iterative-solution}), how words are build up from the letters. In this sense we can think of the constant matrices $a_{i}$ in eq. (\ref{definition-Atilde})  as the `grammar' needed to form words.

It is convenient to introduce a shorthand notation for the iterated integrals. 
For simplicity, let us first give the definition for iterated integrals of a single variable $x$, and choose (here) as boundary point of the integration $x_0 = 1$.
We denote iterated integrals by brackets $[\ldots]$.
Integrals are defined iteratively, namely
\begin{align}
[\alpha_1(x), \ldots , \alpha_{n-1}(x), \alpha_{n}(x)] = \int_0^1 [\alpha_1(x'), \ldots , \alpha_{n-1}(x')] \, d\,\log \left( \alpha_{n}(x') \right) \,.
\end{align}
Here the integration goes along the path $x' =  (1-t) + x t $, with $t \in [0,1]$.
The iteration starts with the empty bracket $[\,] \equiv 1$.
The number of entries $n$ is called weight. 
For example, we have the weight-one function
\begin{align}
[ x ] = \int_0^1 \frac{dt (x-1)}{ (1-t) + x t} =  \log (x) \,,
\end{align}
and the weight two function
\begin{align}
[ x, 1-x] = \int [x'] \, d\log(1-x') =  - {\rm Li}_{2}(1-x) \,.
\end{align}
An important difference of the pentagon functions w.r.t. the single-variable case is that they are defined for the five-dimensional kinematic space $\{ s_{i,i+1} \}$.
Just as in the above examples, the integrals are defined along a path, parametrized by a variable $t\in [0,1]$.
Let us start with a single integral.
We denote
\begin{align}
[ W_{16} ] = \int_\mathcal{C} d \log W_{16} \,,
\end{align}
where integration path $\mathcal{C}$ starts at the boundary point $(-1,-1,-1,-1,-1)$, and goes to the function argument (which we assume to be in the Euclidean region for now), without picking up monodromies.
To be completely explicit, we could choose a straight path $v_{i}(t) = t-1+ t\, v_i$, parametrized by $t \in [0,1]$.
Consequently, we have 
\begin{align}
[ W_{16} ] = \int_0^{1} \frac{dt (1+v_4 - v_1 -v_2 )}{t-1+ t (v_4-v_1-v_2)} = \log (-v_4+ v_1 +v_2 ) \,.
\end{align}
This is valid for $v_i <0, v_1+v_2-v_4>0$. Other kinematic regions can be obtained via analytic continuation.
Just as above, we can define iterated integrals.
There is one new feature of the multi-variable case with respect to  the single-variable one.
In order for the integral to be well-defined, it is important that it is independent of the choice of contour,
as long as singularities are not crossed. In other words, the integral has to be homotopy invariant.
These conditions are called integrability conditions \cite{Goncharov:2010jf}. 
For a generic term of the form $\sum c_{i,j,\vec{a},\vec{b}} [\vec{a} ,W_{i}, W_{j}, \vec{b} ]$, and two variables $x,y  \in \{v_{1},v_{2},v_{3},v_{4},v_{5} \}$, the latter read
\begin{align}
\sum c_{i,j,\vec{a},\vec{b}} [\vec{a} , \vec{b} ]  \, \left( \partial_x \log W_{i} \, \partial_y  \log W_{j} - \partial_y \log W_{i} \, \partial_x  \log W_{j}    \right) = 0 \,.
\end{align}
This equation puts a constraint on the consecutive entries in the iterated integrals.
In section \ref{sec:pentagonfunctions}, we will classify all iterated integrals built from the alphabets 
$\mathbb{A}_{\rm P}$ and $\mathbb{A}_{{\rm P};1}$ satisfying the integrability conditions.

\subsection{Properties of the pentagon alphabet}

\renewcommand{\arraystretch}{1.3}
\begin{table}[t]
\begin{center}
\begin{tabular}{ |c|c|c|c|} %|c
 \hline
   \shortstack{Letter\,\,\,} & \shortstack{$v$ notation\,\,\,} & \shortstack{\ momentum notation \,\,}  &   \shortstack{\ cylic \,\,}  %& 
     \\
 \hline  \hline
% & &    \\
$ W_1 $  & $v_{1}$ &  $2 p_{1} \cdot p_{2} $ & + cyclic (4)   
  \\ 
 $W_{6}$  & $v_{3}+v_{4}$ &  $2 p_{4} \cdot (p_{3}+p_{5}) $ & + cyclic (4)   
  \\ 
$W_{11}$  & $v_{1}-v_{4}$ &  $2 p_{3} \cdot (p_{4}+p_{5}) $ & + cyclic (4) 
  \\ 
$W_{16}$  & $v_{4}-v_{1}-v_{2}$ &  $2 p_{1} \cdot p_{3} $ & + cyclic (4)   
  \\ 
 $W_{21}$  & $v_{3}+v_{4}-v_{1}-v_{2} $ &  $2 p_{3} \cdot (p_{1}+p_{4}) $ & + cyclic (4)   
  \\ 
$W_{26}$  & $\frac{v_1 v_2 -v_2 v_3 +v_3 v_4 -v_1 v_5 -v_4 v_5 - \sqrt{\Delta} }{v_1 v_2 -v_2 v_3 +v_3 v_4 -v_1 v_5 -v_4 v_5 + \sqrt{\Delta}}$ &  $ \frac{ {\rm tr}[(1 -\gamma_{5}) \slashed{p}_{4} \slashed{p}_{5} \slashed{p}_{1} \slashed{p}_{2} ]  }{{\rm tr}[(1 +\gamma_{5}) \slashed{p}_{4} \slashed{p}_{5} \slashed{p}_{1} \slashed{p}_{2} ] } $ & + cyclic (4)    
\\
$W_{31}$  & $\sqrt{\Delta}$ &  $  {\rm tr}[\gamma_{5} \slashed{p}_{1} \slashed{p}_{2} \slashed{p}_{3} \slashed{p}_{4} ] $ &  
\\ \hline 
  \end{tabular}
\end{center}
\caption{Interpretation of pentagon alphabet in terms of particle momenta.}
  \label{tableWletters}
\end{table}
\renewcommand{\arraystretch}{1.0}

The iterated integrals that can appear in our case are characterized by the alphabet $\mathbb{A}_{\rm P}$.
We wrote the alphabet in terms of the Mandelstam invariants $s_{i,i+1}$. 
As we will see, it is instructive to rewrite it in terms of different variables.
This will allow us to see an underlying simplicity of this alphabet.
Let us discuss the different types of letters, and reveal their simple dependence on the external momenta $p_{i}$. This will give us insights into simple parametrizations of the alphabet, and into the singularity structure of the associated functions.

The first $25$ letters $W_1$ to $W_{25}$ (out of which $20$ belong to $\mathbb{A}_{P}$) are simple scalar products of the loop momenta. 
Their interpretation is straightforward: these are possible singularities of Feynman integrals.
One could have discovered these, for example, by an analysis of the Landau equations.
Feynman integrals are multivalued functions, so it is expected that they can have branch cuts. 
However, for planar integrals, only the first five singularities correspond to branch cuts on the first sheet of the functions.
All these letters appear already in four-point integrals with one leg off-shell~\cite{Gehrmann:2000zt,Gehrmann:2001ck}, 
with the planar ones shown in Fig.~\ref{fig:allpentagonintegrals4pt} (and cyclic permutations thereof). 

Next, we have the odd letters $W_{26}\,, \ldots W_{30}$, that are genuine to five-particle kinematics. 
They appear already at one loop, in particular in the six-dimensional pentagon integral.
Remarkably, as pointed out in ref.~\cite{Gehrmann:2015bfy}, they can be written as ratios of traces with a very simple dependence on the momenta, e.g.\
\begin{align}
W_{26} =  \frac{ {\rm tr}[(1 -\gamma_{5}) \slashed{p}_{4} \slashed{p}_{5} \slashed{p}_{1} \slashed{p}_{2} ]  }{{\rm tr}[(1 +\gamma_{5}) \slashed{p}_{4} \slashed{p}_{5} \slashed{p}_{1} \slashed{p}_{2} ] } \,.
\end{align}
Note that writing the letters as ratios is a choice. Using the property of the logarithm 
$d \log(\alpha \beta) = d \log\alpha + d\log \beta$, we could have equally well chosen their numerators as independent letters. 
This is possible since, e.g. 
\begin{align}
 {\rm tr}[(1 -\gamma_{5}) \slashed{p}_{4} \slashed{p}_{5} \slashed{p}_{1} \slashed{p}_{2} ]    {\rm tr}[(1 +\gamma_{5}) \slashed{p}_{4} \slashed{p}_{5} \slashed{p}_{1} \slashed{p}_{2} ] = 4\, v_1 v_4 v_5 (v_5 - v_2 -v_3) \,,
\end{align}
which does not contain new letters.
We prefer to use the ratios, as they have simple transformation properties under parity. 

Finally, we have the Gram determinant as an independent letter. Since $\log(\Delta) =2 \log \sqrt\Delta$ we could equally use $\epsilon(1234)= {\rm tr}[\gamma_{5} \slashed{p}_{1} \slashed{p}_{2} \slashed{p}_{3} \slashed{p}_{4} ]  $.

We summarize the letters of the alphabet, and the equivalent ways of expressing them in Table~\ref{tableWletters}. 
From the equivalent representations of the alphabet letters, we can deduce three remarkable properties of the alphabet.

The first property we comment on is that the alphabet is to a large part determined by the singularity structure of Feynman integrals, i.e. the locations in kinematic space where singularities can occur. This can be analyzed, in principle, via Landau equations. We remark that the alphabet knows not only about singularities on the first sheet of the multi-valued functions, but about {\it all} singularities. In principle, the latter can be rather complicated.
What we observe looking at Table~\ref{tableWletters} is that in our case the allowed singularities follow a simple pattern.
\begin{center}
\noindent\fbox{%
    \parbox{14cm}{%
        {\bf Remarkable property $\# 1$:} the singularities of the pentagon alphabet (\ref{fullpentagonalphabet}) all correspond to exceptional configurations of the external momenta $p_{i} \cdot p_{j} = 0$ or $p_{i} \cdot (p_{j} + p_{k}) =0$, or to restricting them to a lower-dimensional subspace $D<4$, where $\Delta =0$.
    }%
}
\end{center}
\medskip
It is an interesting open question whether these properties will continue to hold at higher loop orders, or whether the alphabet may have to be enlarged in that case. This is relevant for understanding the function space at higher loops, and can have bootstrap applications, see \cite{Henn:2018cdp} for a recent discussion. 

The next property we wish to emphasize is somewhat related to the first one, and concerns the linearity of the alphabet in the momenta.
This has an immediate application.
Sometimes, it can be useful to parametrize the alphabet in a way that is rational in a given variable, and as simple as possible. We remark that the above linearity suggests a way of introducing a variable parametrizing one particular direction of the kinematic space.
The idea is to deform the external momenta, while preserving momentum conservation and the on-shell conditions. This concept is well-known and important in the context of BCFW recursion relations for scattering amplitudes \cite{Britto:2005fq}.
Writing the momenta in terms of spinors, $p_{i} = \lambda_{i} \tilde\lambda_{i}$, one defines for example the deformation
\begin{align}
p_{1} \to p_{1} + z \lambda_1 \tilde{\lambda}_5  \,,\qquad p_{5} \to p_{5} - z \lambda_1 \tilde{\lambda}_5 \,.
\end{align}
In this way, one can study the differential equation along the direction $z$.
It turns out that the alphabet depends linearly on the shift parameter $z$.
The simplicity of our alphabet is also related to a similar parametrization considered in ref. \cite{Papadopoulos:2015jft}.
Summarizing, we have
\begin{center}
\noindent\fbox{%
    \parbox{14cm}{%
        {\bf Remarkable property $\# 2$:} the pentagon alphabet (\ref{fullpentagonalphabet}) can be written in a way that is linear in the external momenta. This allows to introduce BCFW shifts, with a linear dependence on the parameter.  
    }%
}
\end{center}
\medskip
The same comment applies to the extension to non-planar pentagon functions of ref.~\cite{Chicherin:2017dob}.

In fact, it is possible to describe not just a single direction of the alphabet in a rational way, but in fact define a change of variables that completely rationalizes them.
There are various ways of doing this.

One way is using a parametrization suggested by the spinor helicity variables.
In four dimensions, we can write the on-shell momenta in terms of spinors, and use the standard bra-ket notation for spinor invariants. In that language, it is clear that letters $W_{1} \ldots W_{25}$ stay simple. Moreover, we have
\begin{align}
W_{26} = \frac{\langle 45 \rangle [51] \langle 12 \rangle [24] }{[45] \langle 51 \rangle [12] \langle 24 \rangle  }\,, \label{eqW26}\\
W_{31} = - \langle 45 \rangle [51] \langle 12 \rangle [24] + [45] \langle 51 \rangle [12] \langle 24 \rangle \,.
\end{align}
Note that while the spinors manifestly solve the on-shell conditions, they are not independent, due to momentum conservation.
In ref.~\cite{Bern:1993mq}, an independent set of spinor products was used to parametrise the five-particle kinematics. 
Given the fact that all letters are simple rational functions in terms of the spinor brackets, it is not surprising that this leads to a rational version of the alphabet.
We collect the variables, as well as the alphabet in this parametrization in Appendix \ref{app-spinors}.

We remark in passing that eq.~(\ref{eqW26}) shows that the parity-odd letters $W_{26}\,,\ldots W_{30}$ can be interpreted as phases for real momenta (in Minkwoski space), i.e. when $\lambda_{i}$ and $\tilde{\lambda}_{i}$ are related by complex conjugation. Indeed, in that case we have $| W_{26} | =1$.

A parametrization closely related to spinors is that of momentum twistors. The latter variables have the advantage of solving both the on-shell and momentum conservation constraints,
and are hence free variables. Additionally, they offer a geometric interpretation of the singularities of the functions. Like the spinor parametrization, they rationalize the alphabet. We review this in Appendix \ref{app-twistors}.

In summary, we see a third remarkable feature of the pentagon alphabet:
\begin{center}
\noindent\fbox{%
    \parbox{14cm}{%
        {\bf Remarkable property $\# 3$:} the pentagon alphabet can be naturally parametrized in a rational way using spinor or momentum twistor variables.
    }%
}
\end{center}
\medskip
Both property $\#2$ and property $\#3$ imply that, if desired, the pentagon functions can be represented by Goncharov polylogarithms.

In summary, the solutions of eq. (\ref{canonicalDEpentagon}) can be expressed as iterated integrals in the space of {\it{ pentagon functions}} built from the alphabet $\mathbb{A}_{P}$.
In principle, we could proceed with solving the differential equations for the master integrals directly.
We find it instructive, however, to first study the function space from a slightly more general point of view.
This will then allow us to express the solutions of the differential equations in terms of a minimal number of functions.

\section{Classification of planar pentagon functions}
\label{sec:pentagonfunctions}

The differential equations we provided can in principle be solved to any desired order in the $\epsilon$ expansion.
For physical applications to two-loop amplitudes, one typically wishes to evaluate amplitudes up to and including the finite part, but drops $\mathcal{O}(\epsilon)$ contributions.
Although there is no general proof to our knowledge, it is generally believed that two-loop Feynman integrals in four dimensions involve weight four functions at most.
For this reason we focus on this case in particular, and discuss in detail the functions that make an appearance.

\subsection{Planar pentagon functions to weight four}

\begin{table}[t]
\caption{Classification of pentagon functions up to weight four}
\begin{center}
\begin{tabular}{|c|c|c|c|c|}
weight & 1 & 2 &3 &4 \\
\# functions & 5 & 25 & 126 & 651\\
\# products & 0 & 15 & 85 & 480 \\
\# new & 5 & 10 & 41 & 171 \\
no odd letters & 1 $\times$ 5 & 2 $\times$ 5 & 8 $\times$ 5 & 31 $\times$ 5\\
one odd letter & 0 & 0 & 1 & 2 $\times$ 5  +1 \\
two odd letters & 0 & 0 & 0 & 1 $\times$ 5 
\end{tabular}
\end{center}
\label{tableclassification1}
\end{table}%

Physical amplitudes are expected to have branch cuts only originating at $v_{i}=0$.
Therefore we will focus on functions where only $v_{i}$ appear in the first letter.
We proceed by writing all possible words in the alphabet of a given weight,
and imposing this first entry condition, as well as the integrability conditions.

Table \ref{tableclassification1} summarizes the classification of pentagon functions up to weight four.
In the second line, the number of integrable symbols at a given weight is given.

When working with iterated integrals, products of lower-weight integrals can be expanded (via the shuffle algebra)
into sums of higher-weight functions. Inverting these relations, we can remove all product terms from a given expression,
so that only `irreducible' functions remain. This organisation is very useful, as the product terms are faster to evaluate (numerically).
Let $n_1,n_2,n_3,n_4$ be the number of irreducible functions at weight $1,2,3,4$. Then, the number of product functions at weights $2,3,4$ is
\begin{align}
& \#{\rm prod}(2) = \frac{1}{2}(1+n_1) n_1 \,, \\
&  \#{\rm prod}(3) = \frac{1}{6} n_1 (1+n_1) (2+n_1) + n_1 n_2 \,,\\ 
& \#{\rm prod}(4) = \frac{1}{24} n_1 (1+n_1)(2+n_1)(3+n_1) + \frac{1}{2} n_1 (1+n_1) n_2 + \frac{1}{2} n_2 (1+n_2) + n_1 n_3\,,
\end{align} 
respectively.
The number of products functions and of irreducible (new) functions are given in lines 3 and 4 of Table~\ref{tableclassification1}.

The functions can be further classified by the number of entries containing 
the parity-odd letters letters $W_{26}$ to $W_{30}$. This is done in the last three lines 
of Table~\ref{tableclassification1}.
Finally, integrals can be organised according to cyclic symmetry, into either quintets or singlets. For example, at weight one, the five functions all are obtained by cyclic symmetry from one basic function.

At this stage we could proceed and select specific functions representing the various entries on the last three lines 
of Table \ref{tableclassification1}. These would be guaranteed to cover the solution space of the differential equations.
However, we anticipate a further simplification.
As we will see, most functions that are needed actually depend on fewer than the $26$ letters. 
This is closely related to a conjectured second entry condition of ref.~\cite{Chicherin:2017dob}.
It turns out that up to weight four, the letters $v_1 + v_2$ and cyclic only appear
in the slashed box function shown in Fig.~\ref{fig:allpentagonintegrals4pt}(h).

This fact motivates to repeat the above classification of pentagon functions 
for a smaller $21$-letter alphabet, with letters $W_{6}\,, \ldots W_{10}$ removed.
As mentioned earlier, this is the one-loop alphabet $\mathbb{A}_{{\rm P};1}$.
We see that in this case, the number of functions needed is reduced considerably, see Table~\ref{tableclassificationv1v2}.
In the following subsections, we will explicitly construct this basis of iterated integrals.
We introduce all functions needed to describe planar five-particle amplitudes at two loops, up to weight four.
The notation we use is $f_{i,j}^{(k)}$, where $i$ refers to the weight, $j$ is a label, and $k = 1\ldots 5$ refers to different cyclic orderings of the same function. For functions that are singlets of the cyclic group, we simply use the notation $f_{i,j}$.

\begin{table}[t]
\caption{Classification of pentagon functions (without $v_1+v_2$ type letters) up to weight four}
\begin{center}
\begin{tabular}{|c|c|c|c|c|}
weight & 1 & 2 &3 &4 \\
\# functions & 5 & 20 & 76 & 291\\
\# products & 0 & 15 & 60 & 240 \\
\# new & 5 & 5 & 16 & 51 \\
no odd letters & 1 $\times$ 5 & 1 $\times$ 5 & 3 $\times$ 5 & 8 $\times$ 5\\
one odd letter & 0 & 0 & 1 & 1 $\times$ 5  +1 \\
two odd letters & 0 & 0 & 0 & 1 $\times$ 5 \\
total \# needed & 1 & 1  & 4 & 11 
\end{tabular}
\end{center}
\label{tableclassificationv1v2}
\end{table}%

\subsection{Weight one functions}

At weight one, the allowed symbols are dictated by the first entry condition, i.e. the branch cut structure of the integrals,
\begin{align}
 f_{1,1}^{(i)}\,&=\,[\,W_i\,]\;,\;\;\;i\,=\,1\,\dots\,5\; \,.
\end{align}
These integrals evaluate to logarithms.
Using the definition of the iterated integral, with the boundary point $v_{i} =-1$, we have
\begin{align}
 f_{1,1}^{(i)}\,&=\,\log( -  v_i)\;,\;\;\;i\,=\,1\,\dots\,5\;\,.
\end{align}
This formula is manifestly well-defined in the entire Euclidean region $v_{i}<0$. 
To define the function in other regions, one adds the usual Feynman $+i0$ prescription, i.e. $-v_{i} \to -v_{i} - i0$,
and analytically continues~\cite{Gehrmann:2002zr}. 

\subsection{Weight two functions}

It turns out that the combination of first entry conditions and integrability of the iterated integrals is rather restrictive.
Apart from products of weight-one functions, only one new type of function is needed.
It is given by
\begin{align}\label{weight2def}
   f_{2,1}^{(i)}\,&=\,
 \left[
    {W_i \over W_{i+2}}\,,\,
    {W_{i+12} \over W_{i+2}} 
 \right]  \,=\,   
   -{\rm Li}_2 \left(1 - {v_i\over v_{i+2}} \right)\;,\;\;\;i\,=\,1\,\dots\,5 \,.
\end{align}
Note that when generalizing to the non-planar pentagon alphabet $\mathbb{A}_{\rm NP}$, a new type of weight two parity-odd function appears \cite{Chicherin:2017dob}.

At this stage it is crucial to note that the weight one and two functions are very important.
Following \cite{Caron-Huot:2014lda}, all functions up to weight four can be written in terms of one-fold integral representations.
We will discuss this in section \ref{sec:onefold-representations}.
For this reason, in general at higher weight it will be sufficient to define the necessary functions, 
and give the representations we use for their numeric evaluation. 
For simple cases, we can of course use explicit representations in terms of polylogarithms similar to the ones given above.

\subsection{Weight three functions}

At weight three,  according to Table~\ref{tableclassificationv1v2}, we have three even functions, as well as one odd function.

Here, we need for the first time functions that depend on three of the $v_{i}$ at the same time.
This is typical of four-point kinematics, with one off-shell leg. 
For example, for $I_{39}$ shown in Fig.~\ref{fig:allpentagonintegrals4pt}(i),
the momenta are $p_{1}, p_{2}, p_{3}, p_{4}+p_{5}$, such that the relevant invariants are $v_{1}, v_{2}, v_{4}$.

Two of the functions we define are very simple,
\begin{align}
  f_{3,1}^{(i)} \,=&\,
 \left[
    {W_i \over W_{i+2}}\,,\,
    {W_{i+12} \over W_{i+2}}\,,\,
    {W_{i+12} \over W_{i+2}}
 \right] \,,\\
     f_{3,2}^{(i)} \,=&\, 
 \left[
    {W_{i+2} \over W_i}\,,\,
    {W_{i+12} \over W_i}\,,\,
    {W_{i+12} \over W_i}
 \right] \,.
 \end{align}
 These iterated integrals can be done explicitly in terms of polylogarithms.
 For this, we use the definition of the iterated integrals.
 By definition, the first two integrations can be done in terms of the weight one and weight two functions given above.
 We have
 \begin{align}
   f_{3,1}^{(1)} = \int_{\mathcal{C}} f_{2,1,1} d\log \frac{W_{13}}{W_{3}}  = - \int_{\mathcal{C}}    {\rm Li}_2 \left(1 - {v_1 \over v_{3}} \right)  d\log \frac{v_3-v_1}{v_3} \,.
 \end{align}
 In the second equality, we have used eqs. (\ref{weight2def}) and Table~\ref{tableWletters}.
 Here the integration path $\mathcal{C}$ goes from the 
  boundary point $(-1,-1,-1,-1,-1)$ to the function argument (which we assume to be in the Euclidean region for now), without picking up monodromies.
To be completely explicit, we could choose a straight path $v_{i}(t) = t-1+ t\, v_i$, parametrized by $t \in [0,1]$.
Carrying out the integration, we find
\begin{align}
  f_{3,1}^{(i)} \,=&\,
 -{\rm Li}_3 \left(1 - {v_i\over v_{i+2}} \right)\;,\;\label{w3func-Li3}\\
    f_{3,2}^{(i)} \,=&\, 
-{\rm Li}_3 \left(1 - {v_{i+2}\over v_i} \right) \,.
 \end{align}
 These expressions are valid in the full Euclidean region. 
For other regions, one analytically continues, taking into account the Feynman $i0$ prescription.

The third function is slightly more complicated. We define
 \begin{align}
  f_{3,3}^{(1)} \,=&\,\left[{W_1 \over W_4 },{W_{11}\over W_4 },{W_2\over W_4 }\right] \,
   +\,\left[{W_2\over W_4 },{W_{11}\over W_4 },{W_1\over W_4 }\right]\label{w3func-SB}\\
  & -\,\left[{W_1\over W_4 },{W_{11}\over W_4 },{W_{16}\over W_4 }\right]
   \,-\,\left[{W_2\over W_4 },{W_{14}\over W_4 },{W_{16}\over W_4 }\right]\nonumber\\
    & +\,\left[{W_1\over W_4 },{W_2\over W_4 },{W_{16}\over W_4 }\right]
     \,+\,\left[{W_2\over W_4 },{W_1\over W_4 },{W_{16}\over W_4 }\right]\,-\,\zeta_2\,\left[{W_{16}\over W_4 
}\right]\,, \nonumber
\end{align}
with $f_{3,3}^{(i)}$ for $i=2,3,4,5$ obtained from cyclic symmetry.
Here the attentive reader might wonder about the role of the 
$\zeta_2\,  \left[W_{16}\over W_4 \right]$ 
term, especially since $W_{16}$ is not part of the expected branch cuts.
The reason for this term becomes transparent upon explicitly carrying out the first two integrations. We have
\begin{align}\label{f33iterative}
  f_{3,3}^{(1)} \,=& \int_{\mathcal{C}} \Big[ f_{2,1}^{(2)}   d\log \frac{W_1}{W_4}  +
  \left( \frac{1}{2} (f_{1,1}^{(1)})^2 - f_{1,1}^{(1)} f_{1,1}^{(4)} + \frac{1}{2} (f_{1,1}^{(4)})^2 -f_{2,1}^{(4)} \right) 
  d\log \frac{W_2}{W_4}  \\
&  +  \left(-\frac{1}{2}  (f_{1,1}^{(1)})^2   + f_{1,1}^{(1)} f_{1,1}^{(2)} -f_{1,1}^{(2)} f_{1,1}^{(4)} + \frac{1}{2}  (f_{1,1}^{(4)})^2 - f_{2,1}^{(2)} +     f_{2,1}^{(4)}    -\zeta_{2} \right)  d\log \frac{W_{16}}{W_4}  
   \Big] \,.\nonumber 
\end{align}
Recall that the integrals should be well-defined, and be free of branch cuts, in the Euclidean region $v_{i}<0$. One can check that the $\zeta_{2}$ term in the second line of eq. (\ref{f33iterative}) is necessary for the integral to be well-defined. Specifically, it ensures that the function multiplying $d\log W_{16}$  vanishes at $v_{4} = v_{1}+v_{2}$, i.e. when $W_{16}$ vanishes.

Carrying out the integration, we find the following representation
\begin{align}
  f_{3,3}^{(1)} 
  \,=&\,-{\rm Li}_3 {v_1\over v_4} - {\rm Li}_3  {v_2\over v_4}
   + {\rm Li}_3\left( {v_1 + v_2 -v_4\over v_1 v_2} v_4\right)\nonumber \\
 &  - {\rm Li}_3\left({ v_1 + v_2 - v_4 \over v_1}\right)
    - {\rm Li}_3\left( { v_1 + v_2 - v_4\over v_2}\right)   +   3 \* \zeta_3\nonumber \\
 &
 \,+\, \log {v_1\over v_4} \,{\rm Li}_2\left({ v_1 + v_2 - v_4 \over v_2}\right) \,+\,
  \log {v_2\over v_4} \, {\rm Li}_2\left({ v_1 + v_2 - v_4 \over v_1}\right)\label{w3func-new}
\end{align}
This formula for $f_{3,3}^{(1)}$ is valid for the region $v_4<v_1<0, v_4<v_2<0$, a subset of the Euclidean region.
The corresponding formulas for other regions are obtained via analytic continuation. We do not print them here, but they are encoded in an ancillary file. This, and other practical questions about numerical evaluation, are discussed in more detail in sections \ref{sec:boundary} and \ref{sec:numerical-package}.

Finally, there is one parity-odd function at weight three that depends on the full five-point kinematics.
This function can be identified with the (normalized) six-dimensional one-loop pentagon integral $\Phi_{5}$,
\begin{align}\label{defphi5a}
\frac{\Phi_{5}}{\sqrt{\Delta}} = 
 \int_0^{\infty}  \frac{ \prod_{i=1}^{5} dx_{i}  \delta(1-\sum_{i} x_{i})}{[(-v_1) x_1 x_2 + (-v_2) x_2 x_3 + (-v_3) x_3 x_4 + (-v_4) x_4 x_5 + (-v_5) x_5 x_1 ]^2} \,.
\end{align}
We denote it by $f_{3,4}=-\Phi_{5}$.
In our two-loop integral family, it appears as the weight three part of integral $I_{37}$, see Fig.~\ref{fig:genuine5pt}(a).
In terms of iterated integrals, it is given by
 \begin{align}\label{defphi5b}
 f_{3,4} - \frac{2}{3} d_{37,3} 
  \,=\, \int_{\mathcal{C}} 
  \left( \left[ \frac{W_{3}}{W_{5}}, \frac{W_2}{W_{15}} \right] -  
    \left[ \frac{W_{5}}{W_{2}} ,  \frac{W_{3}}{W_{12}}  \right] -\zeta_2 \right)  d\log W_{26}  + {\rm cyclic}\,.
 \end{align}
 Here $-d_{37,3}$ is the value of $\Phi_{5}$ at the symmetric point. Its analytic expression is given in an ancillary file. 
 It follows from the previous subsections that the two inner integrations in the 
iterated-integral definition of $\Phi_5$ can be expressed in terms of logarithms and dilogarithms.
In this way, we obtain the one-dimensional integral representation
  \begin{align} \label{defphi5c}
f_{3,4} - \frac{2}{3} d_{37,3}  \,=\, &\,\,+\,\int_0^1  \,dt\;\partial_t\log\left(
  \frac{a_1(t)+\sqrt{\Delta(t)}}{a_1(t)-\sqrt{\Delta(t)}}
 \right)
 \,\bigg[ \zeta_2  + \log v_1(t) \,\log v_4(t)  \nonumber  \\ &- \log v_4(t)\, \log v_3(t)
  + {1\over2} \log^2 v_3(t) - 
 {1\over2} \log^2v_1(t)\\ &+ {\rm Li}_2\left( 1 - {v_1(t)\over v_3(t)}\right) - {\rm Li}_2\left( 1 - {v_4(t)\over v_1(t)}\right)\bigg]\;+\;{\rm cyclic} \nonumber 
\end{align}
 where the Mandelstam variables $v_i$ are given an implicit $t$ dependence via the parametrization $v_i(t)\,=\,1\,+\,t\,(v_i-1)$.

\subsection{Weight four functions}

We have a total of $9$ parity-even functions without odd letters, one parity-odd function, and one parity-even function with up to two odd letters.
The definitions can be found in an ancillary file.
Here we summarize the main properties of the functions.

Three functions just depend on letters $W_1, W_4, W_{11}$, similarly to functions $f_{3,1}^{(i)}$ and $f_{3,2}^{(i)}$ at weight three.
Five further functions are expressed in terms of the kinematics of the box integrals with one off-shell leg, and depend on $W_{1}, W_{2},W_{4},W_{11},W_{14},W_{16}$ letters only.
As was mentioned earlier, one function corresponds to the slashed box integral $I_{22}$ shown in Fig.~\ref{fig:allpentagonintegrals4pt}(h).
It is the only function containing the letter $W_{9}= v_1 + v_2$.
These functions are sufficient to describe all integrals shown in Figs.~\ref{fig:allpentagonintegralsfactorized} and \ref{fig:allpentagonintegrals4pt}, thereby providing a minimal functional basis for the results of ref.~\cite{Gehrmann:2000zt}.

Finally, in order to describe the integrals of Fig.~\ref{fig:genuine5pt}, only three additional functions are needed.
Let us present their main features (the precise definition is provided in an ancillary file.)
One parity even function is defined as
\begin{align}
 f_{4,11}^{(1)}
    \;=\;  \int_{\mathcal{C}} 
    \left[ \Phi_{5} \, d\log W_{30}   + \ldots \right] \,,
\end{align}
where the dots denote terms required to make the expression integrable.
Similarly, there is a parity odd function
\begin{align*}
  f_{4,10}^{(1)}
  \;=\;  \int_{\mathcal{C}} \left[ \Phi_{5} \,  d\log \left( W_{17} W_{19} W_{31}^4 \right)  + \ldots \right] \,.
\end{align*}
The functions $ f_{4,11}^{(i)}$ and $f_{4,12}^{(i)}$ with $i=2,\ldots 5$ are obtained by cyclic symmetry.
Finally, one parity-odd, cyclically-invariant function reads
\begin{align*}
  f_{4,12}
  \;=\;  \int_{\mathcal{C}} \left[ \Phi_{5} \,  d\log \left( W_{16} W_{17} W_{18} W_{19} W_{20} \right)   + \ldots \right] \,.
\end{align*}
All functions are defined in an ancillary file.
The total number of `irreducible' functions at weights $1,2,3,4$ is $1,1,4,12$, respectively.
The total number of `irreducible' functions is $18$.
The final results for the integrals will be given in terms of these functions.

\begin{center}
\noindent\fbox{%
    \parbox{14cm}{%
        {\bf Summary of pentagon function classification:} All planar pentagon functions up to weight four are expressed in terms of a basis of $18$ irreducible functions, and permutations thereof. Only four of these functions depend on the genuine pentagon kinematics. When expressed in our basis, all identities between functions are manifest. 
    }%
}
\end{center}
\medskip

\subsection{One-fold integral representations for basis functions}
\label{sec:onefold-representations}

In the previous subsections, we defined a complete basis of functions.
As we already mentioned, many functions appeared previously in the context of two-loop massive box integrals with one off-shell leg,
and reliable numerical codes exist for their evaluation, for all relevant kinematic regions \cite{Gehrmann:2001jv}.

Therefore here we only need to discuss the new functions, $f_{3,4}=-\Phi_{5}$ at weight three, and the three functions $  f_{4,10}^{(i)},  f_{4,11}^{(i)},  f_{4,12}$ at weight four.

There are different choices of representations suitable for numerical evaluation. These include expressing the answer in terms of a minimal function basis (consisting of polylogarithms and ${\rm Li}_{2,2}$ functions), in terms of Goncharov polylogarithms, or in terms of an iterated integral representation \cite{Caron-Huot:2014lda}. For a more complete discussion, including examples, we refer the reader to \cite{Caron-Huot:2014lda}, where a similarly complicated alphabet was discussed. The upshot of that discussion is that while the first two types of representations offer advantages, such as a fast and reliable implementation of the basic functions, there are also disadvantages, such as a proliferation of terms. On the other hand, the iterated integrals one starts with are relatively compact. 

What was suggested in \cite{Caron-Huot:2014lda} is a hybrid approach, where one carries out the first two integrations explicitly, in terms the logarithms and dilogarithms discussed above. Naively, it would then seem that one has two integrations remaining to get to weight four. However, by changing the order of integration, another integration can be done explicitly in terms of a logarithm. In this way one gets a one-fold representation that reads, schematically
\begin{align}
f_{4, i}^{(k)} = \int \log(\ldots) {\rm Li}_{2}(\ldots) d \log(\ldots) \,.
\end{align}

 As a concrete example on how the parametrization proposed 
 in ref. \cite{Caron-Huot:2014lda} allowed us to 
 express integrals of a function of given weight
 as integrals of combinations of lower-weight functions
 let us revisit eq. (\ref{defphi5b})  as follows:
  \begin{align}\label{defphi5d}
 f_{3,4} - \frac{2}{3} d_{37,3}    &=
  \int_0^1  \,dt\;\partial_t\log W_{26}(t)
   \left( \left[ \frac{W_{3}}{W_{5}}, \frac{W_2}{W_{15}} \right](t) -  
    \left[ \frac{W_{5}}{W_{2}} ,  \frac{W_{3}}{W_{12}}  \right](t)  \;-\; \zeta_2\right) \;+\;{\rm cyclic} \nonumber \\
  \,=\,&
   \Bigg[ \;
  \int_0^1  \,dt\int_0^{t} \,dt'\;
  \;    \partial_t\log W_{26}(t)\Bigg(
  \; \partial_{t'}\log  \frac{W_2(t')}{W_{15}(t')}
  \; \log \frac{W_{3}(t') }{W_{5}(t') }  
   \; -   \\ &    
  \; \partial_{t'}\log  \frac{W_3(t')}{W_{12}(t')} 
   \; \log \frac{W_{5}(t')}{W_{2}(t')}\Bigg)    
   \;-\;\zeta_2\int_0^1 \,dt\;
   \partial_t\log W_{26}(t)
  \Bigg] \;+\;{\rm cyclic} \;,\nonumber 
\end{align}
 where the dependence on the variable $t$ assigned 
 to the letters $W_i$ and to the iterated-integral functions
 indicates that all kinematic variables $v_i$
 are parametrized as below eq. (\ref{defphi5c}).
If we were to carry out this integration explicitly we would obtain the result presented in eq. (\ref{defphi5c}), 
where dilogarithms appear in the integrand.
Alternatively, we can exchange the order with which we integrate over the variables $t$ and $t'$, obtaining
  \begin{align}\label{defphi5e}
 f_{3,4} - \frac{2}{3} d_{37,3}    &= 
   \Bigg[\;
  \int_0^1  \,dt' \int_{t'}^{1} \,dt\;
  \;    \partial_t\log W_{26}(t)\Bigg(
  \; \partial_{t'}\log  \frac{W_2(t')}{W_{15}(t')}
  \; \log \frac{W_{3}(t') }{W_{5}(t') }  
   \nonumber  \\ &    
  -\; \partial_{t'}\log  \frac{W_3(t')}{W_{12}(t')} 
   \; \log \frac{W_{5}(t')}{W_{2}(t')}\Bigg)  \;-\; \zeta_2\int_0^1 \,dt\;
   \partial_t\log W_{26}(t)
   \Bigg]\;+\;{\rm cyclic} \nonumber \\
     \,=\,&
  \int_0^1 \,dt\; \Bigg[
  \; \left( \log W_{26}(1) - \log W_{26}(t) \right)
  \Bigg(
  \; \partial_{t}\log  \frac{W_2(t)}{W_{15}(t)}
  \; \log \frac{W_{3}(t) }{W_{5}(t) }  
   \;   \\ &    
  - \; \partial_{t}\log  \frac{W_3(t)}{W_{12}(t)} 
   \; \log \frac{W_{5}(t)}{W_{2}(t)}\Bigg) 
    -\;\partial_t\log W_{26}(t) \;\zeta_2\; \Bigg]
    \;+\;{\rm cyclic} \;.\nonumber
\end{align}
Just as in  \cite{Caron-Huot:2014lda}, we coded up these one-fold integral representations and evaluated them numerically.
These routines have been implemented in a public code, as described in section \ref{sec:numerical-package}.

\subsection{Integral basis in terms of pentagon functions}

As a result of the classification of the previous sections, we can express the integrals appearing in the differential equations in terms of that function basis.
Here, we give examples for some of the integrals that have genuine five-particle kinematics.
For $I_{37}$ shown in Fig.~\ref{fig:genuine5pt}(a), we have
\begin{align}
 I_{37} \;=\;&{3\over2}\*f_{3, 4}\;\epsilon^3 \;+\; \\&
 \Bigg[ {\rm bc}_{37, 4} +
     {1\over20}\left( f_{4, 12}- 4 \*f_{4, 10}^{(1)} -4\*f_{4, 10}^{(2)} + 6\*f_{4, 10}^{(3)} + 6\*f_{4, 10}^{(4)} + 6\*f_{4, 10}^{(5)} \right)
   -  \nonumber\\&
      {3\over4}\*f_{3, 4}\*\left(2\*f_{1, 1}^{(1)} + 2\*f_{1, 1}^{(2)}+ 3\*f_{1, 1}^{(3)} + 3\*f_{1, 1}^{(4)} +3\*f_{1, 1}^{(5)}\right)
    \Bigg]\;\epsilon^4 \nonumber  \,,   \label{eq:I37-explicit}
\end{align}
We also show the results for the integrals with maximal number of propagators, Fig.~\ref{fig:genuine5pt}(d),
\begin{align}
 I_{59} \;=\;&{1\over2}\*f_{3, 4}\;\epsilon^3 \;+\; \\&      \label{eq:I59-explicit}
  \Bigg[  {\rm bc}_{59, 4} + {1\over60}\*f_{4, 12} 
  - {7\over30}\*f_{4, 10}^{(1)} - {7\over30}\*f_{4, 10}^{(2)} + {1\over10}\*f_{4, 10}^{(3)} + {13\over30}\*f_{4, 10}^{(4)} + {1\over10}\*f_{4, 10}^{(5)} 
   \nonumber\\ & 
  - {3\over4}\*f_{3, 4}\*\left(f_{1, 1}^{(1)} +\*f_{1, 1}^{(2)} +\*f_{1, 1}^{(3)} +{1\over3}\*f_{1, 1}^{(4)} 
  +f_{1, 1}^{(5)}  \right)      
    \Bigg]\;\epsilon^4  \,,   \nonumber\\
    I_{60} \;=\;& -3 \\&      \label{eq:I60-explicit}
\;+\;\left(f_{1, 1}^{(1)} + f_{1, 1}^{(2)} + 2 f_{1, 1}^{(3)} + 2 f_{1, 1}^{(5)}\right)\;\epsilon\nonumber\\&
\;+\;\left(-2 f_{2, 1}^{(1)} + 2 f_{2, 1}^{(2)} - 2 f_{2, 1}^{(4)} + 2 f_{2, 1}^{(5)}\;+\;\ldots\right)\;\epsilon^2\nonumber\\&
\;+\;\Big(4\*f_{3, 1}^{(1)} + 5\*f_{3, 1}^{(2)} - 8\*f_{3, 1}^{(3)} - f_{3, 1}^{(4)} - f_{3, 2}^{(2)} - 8\*f_{3, 2}^{(3)}  \nonumber\\&\quad
  + 5\*f_{3, 2}^{(4)} + 4\*f_{3, 2}^{(5)} - 2\*f_{3, 3}^{(1)} - 2\*f_{3, 3}^{(3)} -
 2\*f_{3, 3}^{(4)}\;+\;\ldots\Big)\;\epsilon^3 \nonumber\\&
\;+\;\bigg(
-4\*f_{4, 1}^{(1)} + 7\*f_{4, 1}^{(2)} + 11\*f_{4, 1}^{(4)} + 4\*f_{4, 1}^{(5)} + 4\*f_{4, 2}^{(1)} +
 11\*f_{4, 2}^{(2)} + 7\*f_{4, 2}^{(4)} - 4\*f_{4, 2}^{(5)}  \nonumber\\&\quad
 + {37\over2}\*f_{4, 3}^{(1)}
 + {3\over2}\*f_{4, 3}^{(2)} + f_{4, 3}^{(3)}
 - {11\over2}\*f_{4, 3}^{(4)}
 - {31\over2}\*f_{4, 3}^{(5)} +       
 {5\over3}\*f_{4, 4}^{(1)}      
 - {4\over3}\*f_{4, 4}^{(2)}
 + {23\over6}\*f_{4, 4}^{(3)} \nonumber\\&\quad
 + {5\over2}\*f_{4, 4}^{(4)} -      
 2\*f_{4, 4}^{(5)} 
 - {2\over3}\*f_{4, 5}^{(1)} + {10\over3}\*f_{4, 5}^{(2)} 
  - 4\*f_{4, 5}^{(3)} -             
 4\*f_{4, 5}^{(4)} 
 + {2\over3}\*f_{4, 5}^{(5)} + {10\over3}\*f_{4, 6}^{(1)}\nonumber\\&\quad
  - {4\over3}\*f_{4, 6}^{(2)} -                   
{2\over3}\*f_{4, 6}^{(3)}
 - {10\over3}\*f_{4, 6}^{(4)} - {2\over3}\*f_{4, 6}^{(5)}
  - {8\over3}\*f_{4, 7}^{(1)} -                   
 2\*f_{4, 7}^{(2)} + 2\*f_{4, 7}^{(3)} 
 + {2\over3}\*f_{4, 7}^{(4)}\nonumber\\&\quad
  - {8\over3}\*f_{4, 8}^{(1)} +            
 {2\over3}\*f_{4, 8}^{(3)} + 2\*f_{4, 8}^{(4)}
 - 2\*f_{4, 8}^{(5)} + {4\over3}\*f_{4, 11}^{(1)} -      
 {2\over3}\*f_{4, 11}^{(3)} - {2\over3}\*f_{4, 11}^{(4)} \;+\;\ldots\bigg)\;\epsilon^4  \,,
\nonumber\\
 I_{61} \;=\;&-{1\over2}\*f_{3, 4}\;\epsilon^3 \;+\; \\&      \label{eq:I61-explicit}
 \Bigg[   {\rm bc}_{61, 4} + {1\over60}\*f_{4, 12}  -           
   {17\over30}\*f_{4, 10}^{(1)} - {17\over30}\*f_{4, 10}^{(2)} + {1\over10}\*f_{4, 10}^{(3)} + {23\over30}\*f_{4, 10}^{(4)} +     
   {1\over10}\*f_{4, 10}^{(5)}  \nonumber\\ & 
    + {1\over4}\*f_{3, 4}\*\left(f_{1, 1}^{(1)} + f_{1, 1}^{(2)} +    
   3\*f_{1, 1}^{(3)} + 7\*f_{1, 1}^{(4)}
 + 3\*f_{1, 1}^{(5)} \right)\Bigg]\;\epsilon^4\nonumber \,,
\end{align}
where the dots in the expression of $I_{60}$ indicate that, at each order in $\epsilon$,
only the functions with the highest weight have been included, whereas products of lower weight
functions and constants have been omitted; notice that the expressions of $I_{37}$,  $I_{59}$ and $I_{61}$
are  complete.

In the above formula, we have already used the explicit boundary values for the integrals
obtained in the following section \ref{sec:boundary}. Some of the boundary constants appearing at
weight four are rather lengthy, so that we have abbreviated them as ${\rm bc}_{i}$. Their values
are provided in \texttt{pentagox$\_$basis2$\_$bdry$\_$weight4.txt}. The results for all integrals expressed in terms of pentagon functions can be found in \texttt{masters-Walphabet$\_$f.txt}.

\section{Boundary conditions}
\label{sec:boundary}

In section \ref{sec:differentialequations}, we derived a system of first-order differential equations
that the master integrals satisfy, and subsequently identified the basis of functions that can appear in the 
solution to these differential equations. To fully specify the solution, we need to find the boundary conditions to 
the differential equations, i.e.\ the values of all integrals at specific kinematical points. 

One might think that this requires a separate calculation of Feynman integrals. 
However, experience shows that boundary conditions for Feynman integrals can
usually be obtained by the differential equations themselves, together with some
physically motivated constraints. We also find this to be the case here. 
All except one boundary constant are obtained from an analysis of the $\tilde{A}$ matrix in eq. (\ref{definition-Atilde}).
The final boundary constant represents an overall normalization, and is fixed by evaluating one
of the trivial bubble type integrals. The latter can be given, to all orders in $\eps$, in terms of $\Gamma$ functions.

The conditions we will use come from the expectation that the integrals should be non-singular
at several hypersurfaces. We have
\begin{itemize}
\item no branch cuts within the Euclidean region, i.e.\ at $v_{1} - v_{3} = 0$;
\item  no branch cuts in the  $u$-channels: $p_{i} \cdot p_{i+2} =0$;
\item we also expect the integrals to be finite at $\Delta =0$, which corresponds to the external momenta lying 
in a three-dimensional subspace. 
\end{itemize}

The predictive power of these conditions comes from the fact that the differential equations, and hence the matrix
$\tilde{A}$, contain singularities at these locations. This means that the general solution to the equations has
singularities at these locations, while these singularities should be spurious for the actual Feynman integrals 
(which may still contain discontinuities there). 

\subsection{Boundary conditions in the Euclidean region}

The conditions described above give relations between values of the integrals evaluated at different hypersurfaces. It is desirable to have an explicit boundary value for all integrals at the same point.
In the Euclidean region, where $s_{i,i+1} <0$, the symmetric point $s_{i,i+1} = -1$ is a particularly convenient choice.\footnote{The attentive reader might notice that some of the alphabet letters are singular at the symmetric point. However, the latter divergences are spurious. Part of the boundary conditions described above precisely encode this fact.}

In order to compare the boundary values at the symmetric point, we transport them there, using the differential equation. In other words, we integrate the differential equation from some boundary point back to the symmetric point. As the problem is homotopy invariant, one is free to choose a convenient path. Often, this can be done in such a way that the answer is relatively simple. 

Let us give an example of this. We choose the parametrization
\begin{align}
s_{12} = - \frac{x}{(1-x)^2} \,,\quad s_{23} = -1 \,,\quad s_{34} = -1 \,,\quad s_{45} = -1 \,,\quad s_{51} = -1 \,.
\end{align}
Along this path, we find the following reduced alphabet
\begin{align}
\mathbb{B} = \{x+1, x ,x-\frac{1}{2} , x-1, x-2 , 1-3 x+x^2, 1-x+x^2 \}\,.
\end{align}
By this we mean that eq. (\ref{canonicalDEpentagon}) becomes

\begin{align}\label{canonicalDEpentagonx}
\partial_x \vec{I}(x;\epsilon) = \epsilon \,  \sum_{i} b_{i} (\partial_x \log \beta_i)  \, \vec{I}(x;\epsilon) \,,
\end{align} 
with the $b_{i}$ being constant matrixes, and $\beta_{i} \in \mathbb{B}$.

The boundary point $\Delta = 0$ corresponds to $x=-1$.
The symmetric point at $s_{12} = -1$ corresponds to $1-3 x+x^2 =0$. We take the smaller solution 
$x_{0} = \frac{1}{2} (3 - \sqrt{5})$. So, we can use eq. (\ref{canonicalDEpentagonx}) to compute the connection between these two points. Some care is required due to the singular point at $x=0$. For definiteness, we start in the Euclidean region, at $x=x_{0}$. When reaching $x=0$, we analytically continue, using the $i0$ prescription of the Feynman integral. Finally, we integrate from $x=0$ to $x=-1$. The result of these integrations is expressed analytically in terms of Goncharov polylogarithms.

We proceed in a similar way for all boundary points. 
Finally, we fix the overall constant of integration (recall that (\ref{canonicalDEpentagon}) is a homogeneous equation) from a trivial bubble integral (\ref{eq:explicit-subrise-expression}). In this way, all constants of integration are fixed.

Having fixed the boundary condition for all master integrals, in principle we could proceed and obtain a boundary value for each of the physical regions described in section \ref{sec:kin} by analytic continuation.
See section 6 of \cite{Henn:2014lfa} for an example of this in a similar context.
We find it most convenient, however, to repeat the above analysis of consistency equations, directly in the physical regions. This is discussed in the following section.

\subsection{Boundary conditions in the physical regions}

To evaluate the Feynman integrals for physical kinematics, we
integrated the differential-equation system 
separately in each of the ten kinematical regions of Table~\ref{tab:channels}, choosing
as boundary point the value of the integral
in points with a high degree of symmetry located inside
each region (see Table~\ref{tab-bc-phys}).
\begin{table}[t]
{\scriptsize
\begin{center} \begin{tabular}{ |c|c|c|c|c|c|c| } 
 \hline
  & \shortstack{\\Incoming\\momenta\\\,\,} & \shortstack{\\Outgoing\\momenta\\\,\,}  & \shortstack{\\Boundary\\point\\($s_{12},s_{23},s_{34},s_{45},s_{15}$)}  & \shortstack{\\Vanishing\\invariant \\\,\,}  & \shortstack{\\Vanishing\\invariant point\\($s_{12},s_{23},s_{34},s_{45},s_{15}$)} \\ 
 \hline  \hline
 1  & 1,2 & 3,4,5 & & $s_{13}$ & (1, -2/3, 1/3, 1/3, -1/2) \\ 
  &  &  & $\left(1, -{1\over3}, {1\over3}, {1\over3}, -{1\over3}\right)$  & $s_{25}$ & (1, -1/2, 1/3, 1/3, -2/3) \\ 
  & &  &  & $s_{24}$ & (1, -2/3, 1/3, 1/6, -1/3) \\ 
 \hline 
 2  & 5,1 & 2,3,4 & & $s_{25}$ & (-2/3, 1/3, 1/3, -1/2, 1) \\ 
  &  &  & $\left(-{1\over3}, {1\over3}, {1\over3}, -{1\over3}, 1\right)$  & $s_{24}$ & (-1/3, 2/3, 1/3, -1/6, 1) \\ 
  & &  &  & $s_{13}$ & (-2/3, 1/3, 1/6, -1/3, 1) \\ 
 \hline 
 3  & 4,5 & 1,2,3 & & $s_{13}$ & (2/3, 1/3, -1/6, 1, -1/3) \\ 
  &  &  & $\left( {1\over3}, {1\over3}, -{1\over3},1, -{1\over3}\right)$  & $s_{24}$ & (1/6, 1/3, -2/3, 1, -1/3) \\ 
  & &  &  & $s_{25}$ & (1/3, 1/6, -1/3, 1, -2/3) \\ 
 \hline 
 4  & 3,4 & 5,1,2 & & $s_{24}$ & (1/3, -1/2, 1, -1/3, 1/2) \\ 
  &  &  & $\left( {1\over3}, -{1\over3},1, -{1\over3}, {1\over3}\right)$  & $s_{13}$ & (1/3, -2/3, 1, -1/3, 1/6) \\ 
  & &  &  & $s_{25}$ & (1/3, -1/6, 1, -1/3, 2/3) \\ 
 \hline 
 5  & 2,3 & 4,5,1 & & $s_{13}$ & (-1/2, 1, -1/3, 1/2, 1/3) \\ 
  &  &  & $\left( -{1\over3},1, -{1\over3}, {1\over3}, {1\over3}\right)$  & $s_{25}$ & (-2/3, 1, -1/3, 1/6, 1/3) \\ 
  & &  &  & $s_{24}$ & (-1/2, 1, -2/3, 1/3, 1/3) \\ 
 \hline  \hline \hline
 6  & 3,5 & 1,2,4 & & $s_{13}$ & (1/3, -1/2, -1/2, -1/6, -2/3) \\ 
  &  &  & $\left({1\over3}, -{1\over3}, -{1\over3}, -{1\over3}, -{1\over3}\right)$  & $s_{25}$ & (1/3, -2/3, -1/6, -1/2, -1/2) \\ 
  & &  &  & $s_{24}$ & (1/3, -1/6, -1/3, -1/3, -1/2) \\ 
 \hline 
 7  & 1,4 & 2,3,5 & & $s_{13}$ & (-1/2, 1/3, -2/3, -1/6, -1/2) \\ 
  &  &  & $\left(-{1\over3}, {1\over3}, -{1\over3}, -{1\over3}, -{1\over3}\right)$  & $s_{25}$ & (-2/3, 1/3, -1/2, -1/2, -1/6) \\ 
  & &  &  & $s_{24}$ & (-1/6, 1/3, -1/2, -1/3, -1/3) \\ 
 \hline 
 8  & 2,5 & 3,4,1 & & $s_{24}$ & (-1/2, -1/2, 1/3, -2/3, -1/6) \\ 
  &  &  & $\left(-{1\over3},-{1\over3}, {1\over3},  -{1\over3}, -{1\over3}\right)$  & $s_{35}$ & (-1/6, -2/3, 1/3, -1/2, -1/2) \\ 
  & &  &  & $s_{13}$ & (-1/3, -1/6, 1/3, -1/2, -1/3) \\ 
    & &  &  & $s_{14}$ & (-1/3, -1/2, 1/3, -1/6, -1/3) \\ 
 \hline 
  9$^{*}$  & 1,3 & 4,5,2 & & $s_{35}$ & (-1/6, -1/2, -1/2, 1/3, -2/3) \\ 
  &  &  & $\left(-{1\over3}, -{1\over3}, -{1\over3}, {1\over3}, -{1\over3}\right)$  & $s_{14}$ & (-1/2, -1/6, -2/3, 1/3, -1/2)\\ 
  & &  &  & $s_{25}$ & (-1/3, -1/3, -1/2, 1/3, -1/6) \\ 
    & & & & $s_{24}$  &  (-1/3, -1/3, -1/6, 1/3, -1/2)\\ 
 \hline 
 10  & 2,4 & 5,1,3 & & $s_{25}$ & (-1/2, -1/2, -1/6, -2/3, 1/3) \\ 
  &  &  & $\left(-{1\over3}, -{1\over3}, -{1\over3}, -{1\over3}, {1\over3}\right)$  & $s_{14}$ & (-2/3, -1/6, -1/2, -1/2, 1/3) \\ 
  & &  &  & $s_{13}$ & (-1/6, -1/3, -1/3, -1/2, 1/3) \\ 
    & &  &  & $s_{35}$ & (-1/2, -1/3, -1/3, -1/6, 1/3) \\ 
 \hline 
\end{tabular}\end{center}}
 \caption{
 List of boundary points for each of the ten physical regions, together
 with the $s_{i,i+2}=\Delta=0$ points used to obtain contraints
 from spurious-singularity cancellations.
   The ($^{*}$) indicates the physical region where spurious-singularity cancellations 
in the four points reported leave the boundary condition on integral $I_{38}$ unconstrained. 
The value of the integral in this region can be found by analytically continuation from the Euclidean as described above eq.~(\ref{eq:phys-analyticcont-epsilon}).\label{tab-bc-phys}
}
\end{table}

The strategy we follow to evaluate such values
is to exploit the cancellations of spurious singularities
occurring on the boundary of the physical phase space
to set constraints on value of the integrals.
In particular, as the Gram determinant is negative definite for physical kinematics, each physical region is delimited by a hypersurface defined by the equation $\,\Delta = 0\,$.
On the other hand, all  $s_{i,j}$ invariants, including non-adjacent ones, have a definite sign in the physical region, 
therefore 
also the hyperplanes defined by the equations $s_{i,i+2}=0$ delimit the physical region.
Points of tangency between the $\,\Delta=0\,$ hypersurface 
and the $s_{i,i+2}=0$ hyperplanes can therefore be reached
from the boundary points through paths
entirely contained in the physical regions. 
In these points of tangency, the differential-equation system
features non-physical singularities.
By not specifying the boundary values of integrals and
by leaving them as unknown parameters, such divergences
will also affect the integrated results, such that constraints on the boundary values themselves can be obtained by imposing the cancellation of these divergences. 
Considering only one point of tangency for each 
physical region does not provide enough constraints
to be able to determine the boundary values of all integrals.
The path to go from the boundary point (B) to the 
point in which spurious singularities appear (S)
is parametrised as follows\,,
\begin{equation}
    \vec{s}(t) \,=\, \vec{s}_{\rm B} \,+\, (\vec{s}_{\rm S}-\vec{s}_{\rm B})\,t\;,\quad t\in [0,1] \,.
\end{equation}
For all points $\vec{s}_{\rm B}$ and $\vec{s}_{\rm S}$
given in the table above, the resulting reduced alphabet 
contains the square root $\sqrt{-1+t^2}$. 
Such alphabet can be linearised with a change of variable
\begin{equation}
   t\,\rightarrow\,\frac{1+(t'+i(1-t')^2}{2 \,(t'+i(1-t'))}\,,
\end{equation}
which has been chosen so that the integration
path is still parametrized by $t'\in [0,1]$.
For the regions in which the incoming momenta are adjacent,
it is necessary to consider three different
$s_{i,i+2}=\Delta=0$ points. For some of the regions corresponding 
to non-adjacent incoming momenta three points are not enough
to constrain the system and a fourth point needs to be considered.
 
For one region ($s_{13}$ channel) we 
obtained one boundary value (for integral $I_{38}$) by analytically continuing
the solution from the Eulidean region. 
This was done by using the one-fold-integral representation of
the Chen iterated integral expression, 
and by assigning a small imaginary part
to the integration variable as follows
\begin{equation}
   \int d\log\alpha \, f(\{s_{i,i+1}\})\,=\,
      \int_0^1 dt\,\frac{\partial_t\alpha}{\alpha\,-\,i\epsilon} \,
      f\Big(
      \Big\{
      -1+(s_{i,i+1}+1) t+i\epsilon\theta(s_{i,i+1})
      \Big\}\Big)
      \label{eq:phys-analyticcont-epsilon}
\end{equation}
where $\alpha$ represents any letter given in Table \ref{tableWletters}.
The numerical evaluation of this Chen iterated integral is far more time-consuming than the 
evaluation of the pentagon basis functions described in Section~\ref{sec:pentagonfunctions}, which is 
why this method is not used as a default for the implementation. 

Table \ref{tab-bc-phys} summarises, for each of the physical regions, 
the chosen boundary point and the auxiliary $u$-channel-zero points
used to constrain the boundary value. 

\section{Checks on the results}
\label{sec:results}

The two-loop pentagon functions derived in the previous section
 can be combined to yield analytical expressions for 
the full set of planar massless five-point master integrals. Previous results on 
these master integrals concern the  four-point sub-topologies (two-loop four-point functions with one off-shell 
leg~\cite{Gehrmann:2000zt,Gehrmann:2001jv}) and a numerical representation of the five-point
integrals~\cite{Papadopoulos:2015jft}. We checked numerically 
that our pentagon functions reproduce these known master 
integrals, finding machine precision agreement 
 in the Euclidean region as well as in the Minkowskian region in all ten physical scattering channels. 
In the Euclidean region, we have moreover performed a numerical in-depth comparison between our expression in 
terms of Goncharov polylogarithms, and the one-dimensional 
integral representation, described in section \ref{sec:onefold-representations}. Again, machine precision agreement is 
found, with the observation that the one-dimensional integral representation evaluates considerably faster. 

These master integrals appear in the expressions for all-massless two-loop five-point scattering amplitudes. 
Results for these amplitudes in the all-gluon case were first obtained for specific helicity configurations~\cite{Badger:2013gxa,Badger:2015lda,Dunbar:2016aux}, and more recently for all 
general-helicity cases~\cite{Badger:2017jhb,Abreu:2017hqn}. These results expressed the respective 
amplitudes as linear combinations of (numerically or analytically determined) integral coefficients, 
multiplying basis integrals. In these previous works, the five-point basis integrals were then evaluated 
numerically (restricted up to now to  
Euclidean kinematics), using the FIESTA sector decomposition code~\cite{Smirnov:2015mct}.  
By re-expressing these basis integrals in terms of our master integrals, analytical expressions for two-loop 
five-gluon amplitudes can be obtained, such as in \cite{Gehrmann:2015bfy} for the all-plus helicity 
amplitude of~\cite{Badger:2013gxa}. An early check of this result was obtained by an 
independent derivation of this specific amplitude~\cite{Dunbar:2016aux}. 

Working with the authors of~\cite{Badger:2017jhb}, we performed~\cite{LL2018proc} a detailed comparison of all 
individual 
integrals and amplitudes in the Euclidean region, obtaining agreement with~\cite{Badger:2017jhb} on at least six significant 
digits  in each case. 
Our analytical expressions were then also used for the  evaluation of the five-gluon amplitudes 
of~\cite{Badger:2017jhb} in several physical kinematic points~\cite{LL2018proc}, which provides further opportunities for the validation of our pentagon functions. 
The correctness of the two-loop pole structure~\cite{Catani:1998bh,Aybat:2006mz} for all helicity 
configurations provides a check of the integrals up to weight three; moreover, 
the cancellation of 
terms proportional to $(D_s-2)^0$ in the all-but-one-minus  helicity configurations  at order $\epsilon^0$ 
provides a non-trivial consistency
check of the weight-four contributions to the integrals.

Finally, point-wise checks of selected master integrals against numerical results using sector decomposition were performed 
with the FIESTA code~\cite{Smirnov:2015mct} in the Euclidean region and with the SecDec~\cite{Borowka:2015mxa} 
code in different Minkowskian regions. 

Taken together, these checks provide strong support for the correctness of the analytical expressions that we
obtained for the pentagon functions, and for their numerical implementation, whose usage is described below.

\section{Numerical implementation}
\label{sec:numerical-package}

The two-loop planar pentagon functions and the master integrals constructed from them have been 
implemented as {\tt C++} functions, using their one-dimensional integral representation for the numerical evaluation. 
The code requires as prerequisites
 the installation of GiNaC~\cite{Bauer:2000cp} (and its dependencies) 
 and of the GNU Scientific Library GSL~\cite{GSL}. 
The computation of the master integrals is performed in two stages, which 
   correspond to two distinct functions.

  The {\tt C++} function evaluating the linearly independent pentagon
   functions is defined in the file \texttt{evaluate$\_$pentagonfunctions.cpp}:
   \begin{verse}
   \texttt{evaluate$\_$pentagonfunctions(double v[5], double error)}
   \end{verse}
   The arguments of this function are a vector of double precision floating-point numbers containing the 
   five kinematic invariants $v_{1\ldots 5}$ and the relative size of the target error for the numerical
   integration. It is recommended to keep this target error well below machine precision.     
   The function returns a 82-dimensional vector of double-precision complex numbers containing  
   the values of all pentagon functions (including the cyclic permutations of the kinematic invariants). 
  
The {\tt C++} function assembling 
the 61 independent master integrals (in the UT basis) and their 
cyclic permutations from these pentagon functions is defined in the file 
\texttt{evaluate$\_$pentagonintegrals.cpp}: 
   \begin{verse}
   \texttt{evaluate$\_$pentagonintegrals(vector<complex<double> > eval$\_$pen)}
   \end{verse}
   Here, the argument \texttt{eval$\_$pen} 
   of the function is the result of  a preceding  function call to the above \texttt{evaluate$\_$pentagonfunctions}: the
   82-dimensional complex double precision vector of the pentagon functions. 
      The output is a nested vector of dimension $61\times5\times5$ of complex double-precision 
      numbers containing the 
   values of the five cyclic permutations (third index) of the different coefficient of the the expansion in $\epsilon$  (second index)  of the 61 master integrals in the UT basis  \texttt{basis2} (first index).
   
   To illustrate the usage of these functions, and example program \texttt{example-code.cpp} is provided. This 
   program evaluates all pentagon functions and all master integrals for a given kinematic point (either selected from 
  predefined 
   16 example points, or entered manually as list of the $v_{1\ldots 5}$), and writes them to an output file. The target 
   integration error is fixed to $10^{-6}$ in this example code. 

After potentially adjusting the pathnames for GiNaC and GSL in the makefile, the code can be complied by \texttt{make}
 and run by \texttt{./example-code}. 
 
 The code is available as an ancillary file to the arXiv submission of this article. It will be part of a new 
\texttt{www.hepforge.org} repository \texttt{PentagonFunctions}, where also the GPL implementation and
computer algebra expressions for all results from this article will be made  available.

The master-integral basis \texttt{basis2} is normalised such that 
its first element (two-loop massless sunrise with scale $s_{23}$) is
\begin{align}
    I_1 \;=\; (-s_{23})^{-2\,\epsilon} 
       \left(
          -1\;+\;{\pi^2\over6}\epsilon^2\;+\;{32\zeta_3\over3}\epsilon^3\;+\;{19\pi^4\over120}\epsilon^4
       \right) \,.   \label{eq:explicit-subrise-expression}
\end{align}

\section{Conclusion and discussion}
\label{sec:conc}

In this paper, we gave details of the computation of all master integrals of massless, planar five-particle integrals at two loops \cite{Gehrmann:2015bfy}. The integral basis was chosen following the ideas of \cite{Henn:2013pwa}, and as a result, the analytic answers are in a simple and transparent form. In particular, all master integrals are pure functions of uniform transcendental weight in their $\eps$ expansion. They are expressed in terms of iterated integrals, with integration kernels given by the {\it pentagon alphabet} (\ref{fullpentagonalphabet}) that we identified in \cite{Gehrmann:2015bfy}.

We classified all planar pentagon functions up to weight four. They are expressed in terms of a basis of $18$ irreducible functions, and permutations thereof. Only four of these functions depend on the genuine pentagon kinematics. Next, we expressed all master integrals in terms of this basis of functions.

We used properties of the iterated integrals to provide efficient one-fold integral representations for all of the basis functions. For convenience of the user, we coded them in a publicly available program. We validated this code as described in section \ref{sec:results}.

The main focus of the present paper was to provide the full analytic solution for the master integrals, and to provide a code for their fast and reliable numerical evaluation. Beyond this, we remark that the information provided here is very flexible, and can be used, for example, for obtaining asymptotic expansions in any desired limit. We refer the interested reader to ref. \cite{Bruser:2018jnc} for more details in a similar setting.

As a result of our analysis, all integrals needed for planar massless two-loop five-particle scattering amplitudes are available analytically, and can be evaluated numerically using the computer code provided with this publication.

The calculation of planar two-loop five-particle scattering amplitudes is already well underway \cite{Badger:2013gxa,Gehrmann:2015bfy,Badger:2015lda,Dunbar:2016aux,Badger:2017jhb,Abreu:2017hqn}. Our results have already been used in \cite{Badger:2017jhb} for obtaining numerical results. We expect that in the future, they can be used for obtaining full analytic answers for all helicity configurations, extending the initial result of \cite{Gehrmann:2015bfy} for the all-plus helicity configuration.

A very interesting open problem are the non-planar massless five-particle integrals. First results \cite{Chicherin:2017dob,Chicherin:2018ubl,Chicherin:2018wes} suggest a natural extension of the pentagon alphabet obtained in ref. \cite{Gehrmann:2015bfy}. It would be interesting to prove this conjecture, and to compute all master integrals, using the methods employed here.

\section*{Acknowledgements}
TG and JMH thank the KITP at UC Santa Barbara  for hospitality during the program ``LHC Run II and the Precision Frontier'', where part of this work was performed.
This work was supported in part by the Swiss National Science Foundation (SNF) 
under contract  200020-175595, by the PRISMA Cluster of Excellence at Mainz University
and by the European Research Council (ERC)  under grants
 {\it MC@NNLO} (340983) and {\it Novel structures in scattering amplitudes} (725110). 

\appendix

\section{Five-particle parametrizations}

Here we collect useful parametrizations for the five-particle kinematics.

\subsection{Momentum twistor space geometry}
\label{app-twistors}

Momentum twistor variables \cite{Hodges:2009hk} (see also the introductory section of 
\cite{Mason:2009qx}) are convenient variables for on-shell kinematics.\footnote{Usually, these variables are used for planar problems only, especially when loop integrands are discussed. However, as long as only the external kinematics are concerned, they can also be used for non-planar problems.}
They solve both on-shell as well as momentum conservation conditions, and hence
are unconstrained variables.

To define the momentum twistors, one first switches from momenta $p_{i}$ to dual coordinates $x_{i}$, with $x_{i+1}-x_{i} = p_{i}$, and $x_{i+5} \equiv x_{i}$.
To each dual point corresponds a pair of twistors, e.g. $x^{\mu}_{1} \leftrightarrow Z^{[i}_{1} Z^{j]}_{2}$. 
In the present problem, we have five momentum twistors  $Z_{1} \,, \ldots Z_{5}$ describing the on-shell data.

\begin{figure}[t]
{\includegraphics[width=0.6\textwidth]{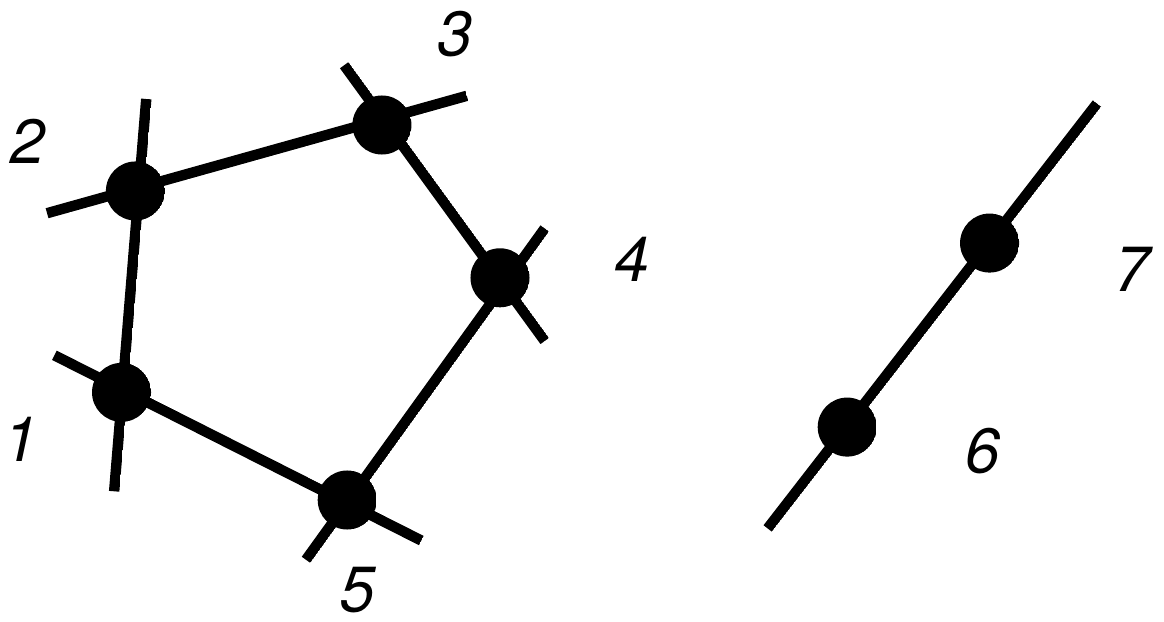}}
\quad\quad\quad
{\includegraphics[width=0.3\textwidth]{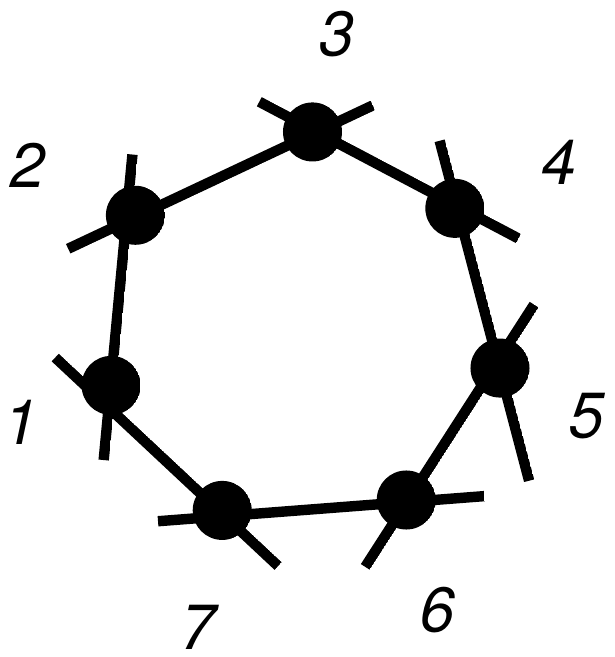}}
\caption{Momentum twistor geometry of (a) five-particle non dual-conformal scattering amplitudes ($5+2$ twistors) and (b) seven-particle scattering with dual conformal symmetry.}
\label{fig:twistorgeometry}
\end{figure}

The $x_{ij}^2$ are simply related to the Mandelstam variables via
\begin{align}
x_{13}^2 = s_{12} \,,\quad 
x_{24}^2 = s_{23} \,,\quad 
x_{35}^2 = s_{34} \,,\quad 
x_{14}^2 = s_{45} \,,\quad 
x_{25}^2 = s_{51} \,. 
\end{align}
To measure distances we also need an ``infinity twistor'', which we write as an auxiliary bitwistor $I =Z_{6} Z_{7}$. With these ingredients at hand, we can write
\begin{align}
x_{13}^2 =& \frac{(5123)}{(5167)(2367)} \,, \quad x_{24}^2 = \frac{(1234)}{(1267)(3467)} 
\,, \quad x_{35}^2 = \frac{(2345)}{(2367)(4567)} \nonumber \\
  x_{14}^2 =& \frac{(3451)}{(3467)(5167)} 
\,, \quad  x_{25}^2 = \frac{(4512)}{(4567)(1267)} 
\,,
\end{align}
where 
\begin{align}
(abcd):= \epsilon_{ijkl}  Z^{i}_a Z^{j}_b Z^{k}_c Z^{l}_d \,.
\end{align}
The light-like conditions are encoded in momentum twistor space geometrically: the corresponding momentum twistor lines intersect,
and we have, e.g. $x_{23}^2 \sim (2334) = 0$. See Fig.~\ref{fig:twistorgeometry}.

It is interesting that there is a closely connected case that can also be described by seven twistors, namely seven-particle scattering in ${\mathcal N}=4$ super Yang-Mills. The latter theory has an additional symmetry, dual conformal symmetry, that implies that the infinity twistor does not appear. So, dual conformal seven-point functions in that theory are described by the data shown in Fig.~\ref{fig:twistorgeometry} (b). It is clear that there are some differences between the two cases, e.g. due to different cyclic symmetry groups. On the other hand, it will be interesting to compare the functions found here to observations about seven-point functions in the $\mathcal{N}=4$ sYM literature.

Of course, the formulation above is a redundant formulation for just five independent
variables $v_{i}$. 
We can choose specific Z's to reduce or remove this redundancy.
In the literature this is sometimes referred as `gauge fixing'.
The only constraint is that the parametrization we choose must be invertible.
We choose\footnote{This is closely related to a parametrization used in \cite{Badger:2013gxa} (See Appendix A.2 there).}
\begin{align}
\left( Z_{1} Z_{2} Z_{3} Z_{4} Z_{5} Z_{6} Z_{7} \right) = \left(
\begin{array}{ccccccc}
 1 & 0 & \frac{1}{x_{1}} & \frac{1}{x_{1} x_{2}}+\frac{1}{x_{1}} & \frac{1}{x_{1}
 x_{2}}+\frac{1}{x_{1} x_{3}}+\frac{1}{x_{1}} & 0 & 0 \\
 0 & 1 & 1 & 1 & 1 & 0 & 0 \\
 0 & 0 & 0 & x_{4} & 1 & 0 & 1 \\
 0 & 0 & 1 & 1 & \frac{x_{5}}{x_{4}} & 1 & 0 \\
\end{array}
\right)\,.
\end{align}
This leads to the following change of variables,
\begin{align}
v_1 &\,=\, x_1\,, \nonumber\\
v_2 &\,=\, x_1 \* x_2 \* x_4\,,\nonumber\\
v_3 &\,=\, \frac{x_1}{x_{2}} [{x_3 \* (x_4 - 1) + x_2\*x_4 + x_2\*x_3\*(x_4 - x_5)}],\\
v_4 &\,=\, x_1 \* x_2 \* (x_4 - x_5),\nonumber\\
v_5 &\,=\, x_1 \* x_3 \* (1 - x_5)\nonumber
\end{align}
The variable $x_1$ is dimensionful, whereas 
the four remaining $\,x_i\,,i=2,\,\dots\,5\,$ are dimensionless.
In other words, $x_{1}$ is a trivial overall scale.
The inverse of this transformation has two branches.
We chose the inversion for which the boundary point (B.P.)
\begin{align} 
 \,v_1=v_2=v_3=v_4=v_5=-1\, \label{mandelstam-sympt}
\end{align}
is given by
\begin{align}
(x_1)_{\rm B.P.} &\;=\; -1\;,\;\;\;\nonumber \\
(x_2)_{\rm B.P.} &\;=\; {1 + \sqrt{5}\over2} \;,\;\;\;\nonumber\\
(x_3)_{\rm B.P.} &\;=\; 1\;,\;\;\; \label{momtwist-sympt}\\
(x_4)_{\rm B.P.} &\;=\; {-1 + \sqrt{5}\over2} \;,\;\;\;\nonumber\\
(x_5)_{\rm B.P.} &\;=\; 0 \,.\nonumber
  \end{align}
This parametrization has the advantage of 
making the Gram determinant a perfect square, leading to
\begin{align}
\sqrt{\Delta}\,=\, -\,x_1^2\*\Big[ x_2\*x_4\*(x_5-1) + x_3\*\Big(1 + x_2\*x_5 
+ x_4\*(-2 - x_2 + x_5)\Big)\Big] \,.
\end{align}
Here, the overall sign was chosen in such a way that $\sqrt{\Delta}$ 
is positive in the symmetric point~(\ref{momtwist-sympt}).

The inverse relations admit two solutions, one of which is
\begin{align}
x_1 &\,\rightarrow\, v_1, \\
x_2 &\,\rightarrow\, \frac{1}{2\*v_1\*v_3}
  \bigg[v_1\*v_2 + v_2\*v_3 - v_3\*v_4 - v_1\*v_5 + v_4\*v_5  +
        \sqrt{\Delta}\bigg],\\
x_3 &\,\rightarrow\,\frac{v_2\*( v_2 - v_4-v_5)}{v_4\*(v_4 - v_1 - v_2)} \nonumber\\& +
        \frac{ v_4 - v_2}{2\*v_1\*v_4\*(v_4 - v_1 - v_2)} \bigg[v_1\*v_2 + v_2\*v_3 - v_3\*v_4 - v_1\*v_5 + v_4\*v_5 
         + \sqrt{\Delta}  \bigg],\\
x_4 &\,\rightarrow\,\frac{1}{2\*v_1\*(v_2 - v_4 - v_5)} \bigg[v_1\*v_2 + v_2\*v_3 - v_3\*v_4 - v_1\*v_5 + v_4\*v_5  -
       \sqrt{\Delta} \bigg],\\
x_5 &\,\rightarrow\, \frac{v_2 - v_4}{2\*v_1\*v_2\*(v_2 - v_4 - v_5)} \bigg[v_1\*v_2 + v_2\*v_3 - v_3\*v_4 - v_1\*v_5 + v_4\*v_5
  -
         \sqrt{\Delta}\bigg]\,.
\end{align}
The second solution is obtained from the former by flipping the sign in front of the square roots.

\subsection{Spinor helicity parametrization}
\label{app-spinors}

In ref. \cite{Bern:1993mq}, a parametrization using spinor helicity variables was employed, see eq. (3) and the text below that equation.
This leads to the parametrization
\begin{align}
s_{12} = -\frac{1}{\alpha_1 \alpha_3}\,,\quad s_{23}= -\frac{1}{\alpha_2 \alpha_4} \,,\quad s_{34} = -\frac{1}{\alpha_3 \alpha_5} \,,\quad s_{45} = -\frac{1}{\alpha_1 \alpha_4} \,,\quad s_{51} = -\frac{1}{\alpha_2 \alpha_5} \,.
\end{align}
Further setting
\begin{align}
\alpha_1 =& -\beta_1 - \beta_2^{\star} \,, \\
\alpha_2 =& -\beta_2^\star - \beta_3 \,,\\
\alpha_3 =& -\beta_3-\beta_4^\star \,,\\
\alpha_4 =& -\beta_4^\star - \beta_5 \,,\\
\alpha_5 =& 1/\beta_3 (\beta_1 \beta_4^\star + \beta_2^\star \beta_4^\star + \beta_2^\star \beta_5)\,,
\end{align}
we have 
\begin{align}
\sqrt{\Delta} = \frac{\beta_1 \beta_3 + \beta_1 \beta_4^\star + \beta_2^\star \beta_4^\star + \beta_2^\star \beta_5 + \beta_3 \beta_5 }{(\beta_1 + \beta_2^\star ) (\beta_2^\star + \beta_3 )(\beta_3 + \beta_4^\star ) (\beta_4^\star+\beta_5 )(\beta_1 \beta_4^\star+ \beta_2^\star \beta_4^\star + \beta_2^\star \beta_5) }\,.
\end{align}

\subsection{Scattering equations parametrization}
\label{app-scattering-eqs}

At five particles, the scattering equations \cite{Cachazo:2013hca} have two complex solutions \cite{Weinzierl:2014vwa} that one can use to parametrize the four scale-invariant ratios appearing at five particles. This leads to the following parametrization\footnote{We thank Claude Duhr for sharing this parametrization with us.}
\begin{align}
v_1 =& \rho \frac{({s}_{1} - \overline{s}_{1}) ({s}_{1} - s_{2}) (\overline{s}_{1} - \overline{s}_{2}) (s_{2} - \overline{s}_{2})}{(-\overline{s}_{1} s_{2} + 
   {s}_{1} \overline{s}_{2}) (-{s}_{1} + \overline{s}_{1} + s_{2} - \overline{s}_{1} s_{2} - \overline{s}_{2} + {s}_{1} \overline{s}_{2})}  \,\\
v_2 =&    - \rho \frac{({s}_{1} - \overline{s}_{1}) s_{2} \overline{s}_{2}}{(-\overline{s}_{1} s_{2} + {s}_{1} \overline{s}_{2})} \,\\
v_3 =& \rho \, \\
v_4 =& \rho \frac{(-\overline{s}_{1} s_{2} + {s}_{1} \overline{s}_{1} s_{2} + {s}_{1} \overline{s}_{2} - {s}_{1} \overline{s}_{1} \overline{s}_{2} - {s}_{1} s_{2} \overline{s}_{2} + 
 \overline{s}_{1} s_{2} \overline{s}_{2})}{(-{s}_{1} + \overline{s}_{1} + s_{2} - \overline{s}_{1} s_{2} - \overline{s}_{2} + {s}_{1} \overline{s}_{2})} \,\\
 v_5 =&   - \rho \frac{ (s_{2} - \overline{s}_{2}) (\overline{s}_{1} s_{2} - {s}_{1} \overline{s}_{1} s_{2} - {s}_{1} \overline{s}_{2} + {s}_{1} \overline{s}_{1} \overline{s}_{2} + {s}_{1} s_{2} \overline{s}_{2} - 
    \overline{s}_{1} s_{2} \overline{s}_{2}) }{(\overline{s}_{1} s_{2} - {s}_{1} \overline{s}_{2}) ({s}_{1} - \overline{s}_{1} - s_{2} + \overline{s}_{1} s_{2} + \overline{s}_{2} - 
    {s}_{1} \overline{s}_{2})}  \,.
    \end{align}
Like in the other cases, these variables allow one to rationalize the square root $\sqrt{\Delta}$.

%\bibliographystyle{JHEP}
%\bibliography{refs}

\providecommand{\href}[2]{#2}\begingroup\raggedright\endgroup

\end{document}